%% file: maintex.tex
\documentclass[a4paper,oneside,12pt]{article} 

\usepackage{axodraw4j}
\usepackage{pstricks}
\usepackage{color}

\usepackage{units}
\usepackage{graphicx}
\usepackage{dsfont}
\usepackage{graphics}
\usepackage{epsfig}
\usepackage{flafter}
\usepackage{amsmath,amssymb,amsfonts,latexsym}
\usepackage{array}
\usepackage{multirow} 
\usepackage{lineno}   
\usepackage{rotating} 
\usepackage{subfigure} 
\usepackage{url}

\begin{document}

\include{title}
\include{prelude}

\include{abstract}
\include{index}
\pagenumbering{arabic}

\include{DiffractionPomeron}

\include{DiffractionPythia}

\include{conclusions}
\include{acknowledgements}


\include{bibliography}
\end{document}

%% file: title.tex
\newcommand{\ia}{\'{i}}

\begin{titlepage}
\begin{flushright}
  LUTP-09-23\\
  MCnet/10/09\\
\end{flushright}

\vspace{30mm}

\begin{center}
  \vspace{1.0cm}
  {\Huge\bf  Diffraction in PYTHIA\\}
\end{center}

\vspace{30mm}

\begin{center}
  {\Large\bf Sparsh Navin}\\
  School of Physics and Astronomy, The University of Birmingham, \\
  Birmingham, United Kingdom
\end{center}
\end{titlepage}

%% file: prelude.tex
\pagestyle{plain}
\pagenumbering{roman}

%% file: abstract.tex
\begin{abstract}
	\addcontentsline{toc}{part}{Abstract}
\noindent The PYTHIA program can be used to generate high-energy-physics ‘events’ with sets of outgoing particles produced in the interactions between two incoming particles. The objective is to provide a representation, as accurate as possible, of event properties in a wide range of reactions. One such reaction, that is not well understood is Diffraction. Among the several alternative approaches that have been proposed, in PYTHIA, we follow a fairly conventional Pomeron based one, but fully integrated to use the standard PYTHIA machinery for multiple interactions, parton showers and hadronization. 
	This note reports the development in PYTHIA in the way diffraction is modeled without providing specific details for usage. Results are compared with an alternative event generator called PHOJET. 
	The code and further information may be found on the Pythia web page: \url{http://home.thep.lu.se/~torbjorn/Pythia.html}
\end{abstract}

%% file: index.tex
\topmargin=3mm \headheight=0mm \headsep=0mm
\tableofcontents

%% file: DiffractionPomeron.tex

\section{Diffraction and the Pomeron}

Hadronic processes can be classified as being either soft or hard. Soft
processes, that dominate hadronic scattering cross-sections, are 
characterised by an energy scale of the order of the hadron size (~\unit[1]{f
m} $\approx$ \unit[200]{MeV}) \cite{barone}. The hard sector is described very well by Perturbative QCD (pQCD) scatterings. However, pQCD is inadequate to describe soft processes, as a large scale makes the coupling constant
($\alpha_s$) large enough to make the higher order terms non-negligible, thus making the process intrinsically non-perturbative. \\
		
\noindent I.~Pomeranchuk predicted that if the total cross-section
behaves asymptotically like a power of $\ln{s}$, then the particle and anti-particle cross sections become asymptotically equal \cite{pomeron}. The exchange of a Regge
trajectory that ensures this behaviour was first introduced by Gribov
\cite{gribov}. The particles on this trajectory are virtual and have the same internal quantum numbers as the vacuum. The effective summation of particles on this trajectory is known as the Pomeron ($\mathds{P}$).
In QCD, the Pomeron is regarded as a
colourless and flavourless multiple gluon \cite{ross} or a glueball exchange. 

\subsection{Classification}
			  
\begin{figure}[htbp]
  \begin{center}
  \input{plots/DiffPom/proc_el_fig}  
	 \caption{\footnotesize{Diagram for elastic scattering and $\phi$ vs $\eta$ plot showing the distribution of products after the interaction.}}
    \label{fig:el_feynman}
  \end{center}
\end{figure}
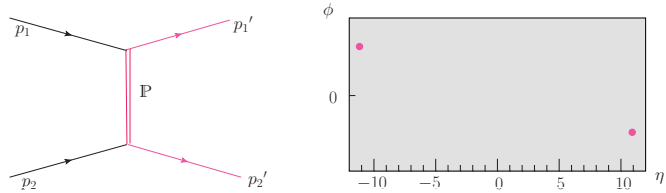

\noindent In proton-proton (pp) (or more generally hadron-hadron) scattering,
interactions are classified by the characteristics of the final
states. Interactions can either be elastic or inelastic. In elastic scattering
($p_1 + p_2 \rightarrow p_1^{\prime} + p_2^{\prime}$), both protons emerge
intact and no other particles are produced as shown by the pink dots in figure
\ref{fig:el_feynman}. The LHC cross-section (at $\sqrt{s}=\unit[14]{TeV}$) for elastic scattering is
$\sim$\unit[30]{mb} \cite{mario}.\\

\noindent Colliding hadrons are colour singlets. As they approach each other, they may
exchange a colour octet gluon, making each hadronic cluster a colour octet. As they
move apart, they need colour lines connecting them. To be able to
separate into two separate systems, they need to exchange another gluon and become colourless. However, the final state need not be identical to the initial state. Such processes are called inelastic. When colour octets move apart, colour lines are stretched between them. Given time, this system gets complex and multi-particle production occurs.\\ 

\noindent Elastic scattering can be achieved via the exchange of a glueball-like
Pomeron. In elastic scattering, the final state and initial state particles are identical. 
The exchange of gluons can excite a hadron. This can result in the
outgoing state preserving the internal quantum numbers of the incoming particles but having a
higher mass. This is known as quasi-elastic scattering.\\

\noindent Inelastic collisions can be diffractive. There are several possible descriptions of diffraction, allowing several alternative approaches. The approach discussed here is one described by Regge theory
\cite{collins} in terms of the exchange of a Pomeron. One of the alternative approaches which does not use the concept of a Pomeron or Regge phenomenology is called the soft colour interaction model. It is described by Ingelman in \cite{gunnaringelman}. \\

\begin{figure}[htbp]
  \begin{center}
  \input{plots/DiffPom/proc_sd_fig}  
    \caption{\footnotesize{SD diagram and a window showing a rapidity gap between $-10<\eta<3.5$.}}
    \label{fig:sd_feynman}
  \end{center}
\end{figure}
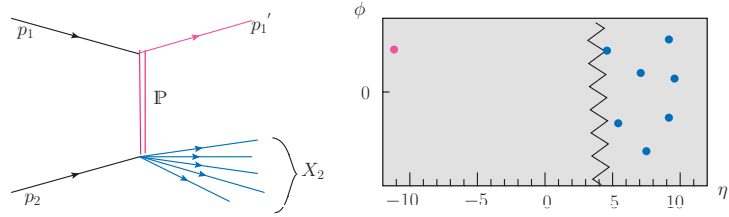

\noindent A diffractive reaction is one in which 
no internal quantum numbers are exchanged between the colliding particles.
Diffraction occurs when the exchanged Pomeron interacts with the proton to
produce a system of particles referred to as the diffractive system.
In diffractive scattering, the energy transfer between the two interacting
protons remains small, but one or both protons dissociate into multi-particle
final states with the same internal quantum numbers of the colliding protons.
\\
If only one of the protons dissociates then the interaction is Single Diffractive (SD)
($p_1 + p_2 \rightarrow p_1^{\prime} + X_2$ or $p_1 + p_2 \rightarrow X_1 +
p_2^{\prime}$). The dissociated proton is shown as a spray of blue dots
(particles) and the non-dissociated proton as the pink dot in figure
\ref{fig:sd_feynman}. The LHC cross-section (at $\sqrt{s}=\unit[14]{TeV}$) for SD is $\sim\unit[10]{mb}$
\cite{mario}. \\

\begin{figure}[htbp]
  \begin{center}
  \input{plots/DiffPom/proc_dd_fig}  
    \caption{\footnotesize{DD diagram and window showing a rapidity gap between $-3.5<\eta<4$.}}
    \label{fig:dd_feynman}
  \end{center}
\end{figure}
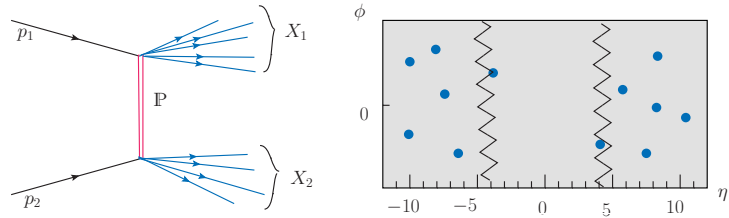

\noindent If both the colliding protons dissociate, then it is Double
Diffractive (DD) ($p_1 + p_2 \rightarrow X_1 + X_2$) as seen in figure
\ref{fig:dd_feynman}. The LHC cross-section (at $\sqrt{s}=\unit[14]{TeV}$) for DD is $\sim\unit[7]{mb}$
\cite{mario}.\\

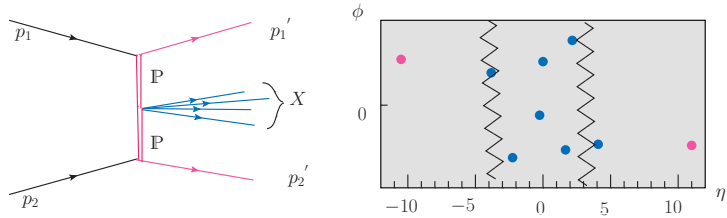
\begin{figure}[htbp]
  \begin{center}
  \input{plots/DiffPom/proc_cd_fig}  
    \caption{\footnotesize{CD diagram and window showing two rapidity gaps between $-10<\eta<-2.5$ and $2.5<\eta<10$.}}
    \label{fig:cd_feynman}
  \end{center}
\end{figure}

\noindent A different topology becomes possible with two Pomerons exchanged, namely Central Diffraction (CD)
($p_1 + p_2 \rightarrow p_1^{\prime} + X + p_2^{\prime}$) or Double Pomeron
Exchange. In this process, both the protons are intact and are seen in the
final state (as two pink dots seen in figure \ref{fig:cd_feynman}). The LHC
cross-section for CD is $\sim\unit[1]{mb}$ \cite{mario}.\\

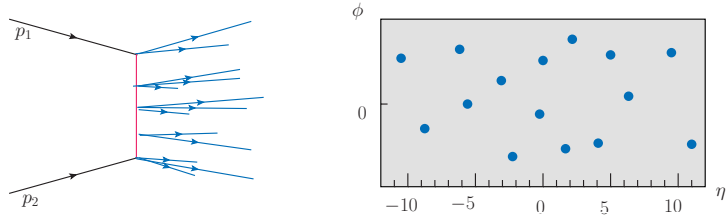
\begin{figure}[htbp]
  \begin{center}
  \input{plots/DiffPom/proc_nd_fig}  
    \caption{\footnotesize{The diagram for an ND process. The rapidity window on the right shows that there is no rapidity gap.}}
    \label{fig:nd_feynman}
  \end{center}
\end{figure}

\noindent In Non-Diffractive (ND) interactions there is an exchange of colour charge and subsequently more
hadrons are produced. This is shown in figure \ref{fig:nd_feynman}. ND
interactions are the dominant process in pp interactions and are expected to be
$\sim$60\% of all interactions at the LHC with a cross-section of
$\sim$\unit[65]{mb} (at $\sqrt{s}=\unit[14]{TeV}$) \cite{mario}.\\

\noindent A consequence of the Pomeron hypothesis is that the cross-sections
of pp and p$\bar{\mathrm{p}}$ diffractive scattering should be equal at high
enough energies \cite{landshoff}. This is because the Pomeron has the quantum
numbers of the vacuum, so its couplings to the proton and anti-proton are
equal.

\noindent The total pp cross-section is given by equation \ref{E:cs} where ``$\textrm{misc}$" here is CD and multiple Pomeron exchange. The cross-section for multiple Pomeron exchange is $\ll$~\unit[1]{mb} \cite{mario}.

\begin{equation}\label{E:cs}
  \sigma_{\textrm{tot}}=\sigma_{\textrm{el}}+\sigma_{\textrm{inel}}=\sigma_{\textrm{el}}+\sigma_{\textrm{diff}}+\sigma_{\textrm{ND}}=\sigma_{\textrm{el}}+\sigma_{\textrm{SD}}+\sigma_{\textrm{DD}} +\sigma_{\textrm{misc}}+\sigma_{\textrm{ND}}
\end{equation}

\subsection{Kinematics}

\noindent In a QCD approach, a partonic description of a Pomeron, as described in \cite{lownussinov} is commenly used. Distributions of partons in particles are
characterised by Parton Distribution Functions (PDF). A PDF $f_{i}(x,Q^{2})$
gives the probability of finding a parton \textit{i} with a fraction \textit{x} of the momentum of the parent beam particle, when probed at a
scale of $Q^{2}$. PDFs are parameterisations of experimental data. 
Diffractive hard scattering is used to resolve the partonic structure of the Pomeron \cite{UA8HERA}. 
 
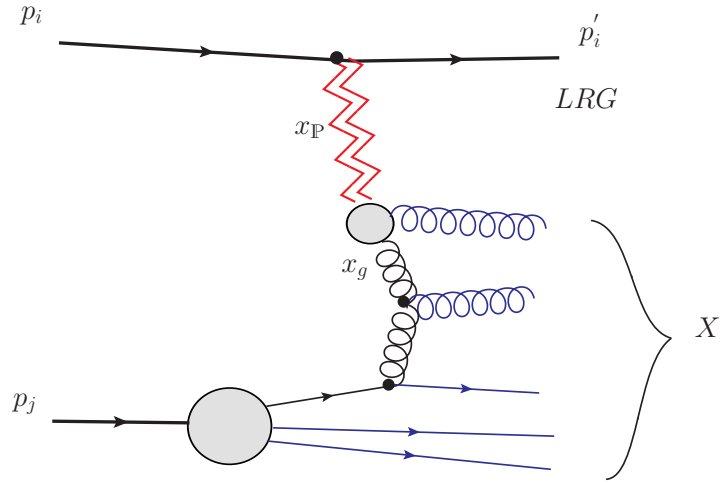
\begin{figure}[htbp]
  \begin{center}
   \input{plots/DiffPom/SD} 
    \caption{\footnotesize{Exchange diagram for single diffraction.}}
    \label{fig:SD_jaxodraw}
  \end{center}
\end{figure}

\noindent Different alternative factorizations of the partonic structure of the Pomeron exist. 
A model for diffractive hard scattering is described in \cite{ingelmanschlein}. 
In this type of factorisation, firstly a Pomeron is emitted from a proton $p_i$ (at the upper vertex in figure \ref{fig:SD_jaxodraw}) with a
momentum transfer squared given by

\begin{equation}\label{E:t}
  t=(p_i-p_i^{\prime})^2. 
\end{equation}

\noindent Then this emitted Pomeron interacts with the
other proton, $p_j$ at the lower vertex, with a transfer of momentum between
constituent partons. The system \textit{X} that is produced in this interaction is called the diffractive system. There is a large rapidity gap (LRG) between the out-going proton and diffractive system \textit{X}. This introduces the concept of a Pomeron flux in a proton (\textit{p})
$f_{\mathds{P}/p}(x_{\mathds{P}},t)$ (in this case $f_{\mathds{P}/p_i}(x_{\mathds{P}},t)$), where $x_{\mathds{P}}$ is the fraction
of the proton's momentum carried by the emitted Pomeron, and diffractive
PDFs (DPDF). The Pomeron flux describes the probability that a Pomeron with a given value
of $x_{\mathds{P}}$ and \textit{t} couples to the proton. 

\noindent In the massless limit,
\begin{equation}\label{E:xp}
  x_{\mathds{P}}=E_{\mathds{P}}/E_{p}
\end{equation}
\\
\noindent where $E_{\mathds{P}}$ and $E_p$ are the energy of the Pomeron and the  proton to which it was coupled to respectively. The fraction of the Pomeron's momentum carried by its constituent gluon (g) (or quark (q)) is given by 

\begin{equation}\label{E:xg}
  x_{\textrm{g (or q)}}=E_{\textrm{g (or q)}}/E_{\mathds{P}}
\end{equation}
\\		
\noindent where $E$ is the energy of the gluon (or quark).

\noindent The diffractive hard pp scattering cross-section can be written as

\begin{equation}\label{E:diffcs}
  \frac{d\sigma(pp\rightarrow p+X)}{dx_{\mathds{P}}dtdx_1dx_2d\hat{t}}=\underbrace{f_{\mathds{P}/p}(x_{\mathds{P}},t)}_{\mathds{P}\textrm{flux}}\frac{d\sigma(p\mathds{P}\rightarrow X)}{dx_1dx_2d\hat{t}}.
\end{equation}

\vspace{10mm}

\noindent The second term in equation \ref{E:diffcs} is the proton-Pomeron hard scattering
differential cross-section. It can be assumed to factorize as in equation \ref{E:pPcs}:

\begin{equation}\label{E:pPcs}
  \frac{d\sigma(p\mathds{P}\rightarrow X)}{dx_1dx_2d\hat{t}}=f_{p_1/p}(x_1,Q^2)f_{p_2/\mathds{P}}(x_2,Q^2)\frac{d\hat{\sigma}}{d\hat{t}}.
\end{equation}

\vspace{10mm}

\noindent Here, $f_{p_1/p}(x_1,Q^2)$ and $f_{p_2/\mathds{P}}(x_2,Q^2)$ are the proton and
Pomeron PDFs with partons $p_1$ and $p_2$ having momentum fractions $x_1$ and $x_2$
of the proton and Pomeron respectively. $d\hat{\sigma}/d\hat{t}$ is the
corresponding hard scattering cross-section for that subprocess. Because of the inherent non-perturbative effect in a QCD binding state, PDFs (and DPDFs) cannot be obtained by perturbative QCD from first principles. The known PDFs (and DPDFs) are instead obtained by using fits to experimental data. The DPDFs used here are obtained using the DGLAP evolution equations \cite{DGLAP}. The
invariant mass of the diffractive system $X$, also known as the diffractive mass, is given in terms of the overall
collision cms energy $\sqrt{s}$ by

\begin{equation}\label{E:Mx}
  M_X^2=x_{\mathds{P}}s.
\end{equation}
\\
\noindent Experimentally, diffractive reactions are characterised by a large
(non exponentially-suppressed) rapidity gap in the forward region, e.g, $x_{\mathds{P}}\leq0.1$.
In other words, there is a large separation in rapidity between the
quasi-elastically scattered proton and the diffractive system, in which no
particles are detected. A few ND events may also display a large rapidity gap
due to multiplicity fluctuations but their number is exponentially suppressed
with increasing rapidity gap. 

\noindent Another exchange mediator
called the Reggeon $\mathds{R}$ \cite{reggeon} is needed to  reproduce experimental data of
diffractive processes and total cross-sections successfully. Thus, there are two exchange mediators:
Reggeons and Pomerons. Reggeon exchange fits data at relatively lower energies (high $x_{\mathds{R}}$), as Reggeons couple to valence quarks of a proton, which carry a large fraction of the Pomeron's momentum $x$. At high energies, the incoming protons ``pass by" so quickly that it is mainly the sea quarks that interact.
On the other hand, Pomeron exchange fits the data only at higher energies (low $x_{\mathds{P}}$), as a Pomeron couples to gluons (and sea quarks). Already at ISR energies ($\sqrt{s}=\unit{63}[GeV]$), the exchange mediator was predominently the Pomeron.
Thus, the higher the collision energy, the more important is the role of the Pomeron. The sum of these two trajectories describe the total pp cross-section.\\

\noindent Based on these theories about the Pomeron, a model for diffraction
has been constructed and implemented in Pythia \cite{pythia8.1}, resulting in a
complete final state.

\bigskip

%% file: plots/DiffPom/proc_el_fig.tex


\resizebox{4in}{!}{
\fcolorbox{white}{white}{
  \begin{picture}(731,259) (0,0)
    \SetWidth{1.0}
    \SetColor{Black}
    \Line[arrow,arrowpos=0.5,arrowlength=5,arrowwidth=2,arrowinset=0.2](66,163)(180,131)
    \Line[arrow,arrowpos=0.5,arrowlength=5,arrowwidth=2,arrowinset=0.2](66,7)(180,39)
    \SetColor{WildStrawberry}
    \Line(179,132)(180,39)
    \SetColor{Black}
    \GBox(396,13)(684,163){0.882}
    \Text(372,163)[lb]{\Large{\Black{$\phi$}}}
    \Text(696,1)[lb]{\Large{\Black{$\eta$}}}
    \Line(396,87)(402,87)
    \Line(540,13)(540,25)
    \Line(600,13)(600,25)
    \Line(660,13)(660,25)
    \Line(480,13)(480,25)
    \Line(420,13)(420,25)
    \Line(552,13)(552,19)
    \Line(564,13)(564,19)
    \Line(576,13)(576,19)
    \Line(588,13)(588,19)
    \Line(612,13)(612,19)
    \Line(624,13)(624,19)
    \Line(636,13)(636,19)
    \Line(648,13)(648,19)
    \Line(672,13)(672,19)
    \Line(528,13)(528,19)
    \Line(516,13)(516,19)
    \Line(504,13)(504,19)
    \Line(492,13)(492,19)
    \Line(468,13)(468,19)
    \Line(456,13)(456,19)
    \Line(444,13)(444,19)
    \Line(432,13)(432,19)
    \Line(408,13)(408,19)
    \Text(540,-5)[lb]{\Large{\Black{$0$}}}
    \Text(600,-5)[lb]{\Large{\Black{$5$}}}
    \Text(657,-5)[lb]{\Large{\Black{$10$}}}
    \Text(378,80)[lb]{\Large{\Black{$0$}}}
    \Text(465,-5)[lb]{\Large{\Black{$-5$}}}
    \Text(405,-5)[lb]{\Large{\Black{$-10$}}}
    \Text(72,145)[lb]{\Large{\Black{$p_1$}}}
    \Text(78,-5)[lb]{\Large{\Black{$p_2$}}}
    \SetColor{WildStrawberry}
    \Line (183,132)(183,40)
    \SetColor{Rhodamine}
    \Line[arrow,arrowpos=0.5,arrowlength=5,arrowwidth=2,arrowinset=0.2](180,132)(280,161)
    \Line[arrow,arrowpos=0.5,arrowlength=5,arrowwidth=2,arrowinset=0.2](181,39)(291,7)
    \Text(193,86)[lb]{\Large{\Black{$\mathds{P}$}}}
    \Text(285,149)[lb]{\Large{\Black{${p_1}'$}}}
    \Text(299,-3)[lb]{\Large{\Black{${p_2}'$}}}
    \Vertex(406,135){3.606}
    \Vertex(671,51){3.606}
  \end{picture}
}
}

%% file: plots/DiffPom/proc_sd_fig.tex
\resizebox{4in}{!}{
\fcolorbox{white}{white}{
  \begin{picture}(666,186) (65,-75)
    \SetWidth{1.0}
    \SetColor{Black}
    \Line[arrow,arrowpos=0.5,arrowlength=5,arrowwidth=2,arrowinset=0.2](66,90)(180,58)
    \Line[arrow,arrowpos=0.5,arrowlength=5,arrowwidth=2,arrowinset=0.2](66,-66)(180,-34)
    \SetColor{WildStrawberry}
    \Line(180,61)(181,-32)
    \SetColor{Black}
    \GBox(396,-60)(684,90){0.882}
    \Text(372,90)[lb]{\Large{\Black{$\phi$}}}
    \Text(696,-72)[lb]{\Large{\Black{$\eta$}}}
    \Line(396,24)(402,24)
    \Line(540,-60)(540,-48)
    \Line(600,-60)(600,-48)
    \Line(660,-60)(660,-48)
    \Line(480,-60)(480,-48)
    \Line(420,-60)(420,-48)
    \Line(552,-60)(552,-54)
    \Line(564,-60)(564,-54)
    \Line(576,-60)(576,-54)
    \Line(588,-60)(588,-54)
    \Line(612,-60)(612,-54)
    \Line(624,-60)(624,-54)
    \Line(636,-60)(636,-54)
    \Line(648,-60)(648,-54)
    \Line(672,-60)(672,-54)
    \Line(528,-60)(528,-54)
    \Line(516,-60)(516,-54)
    \Line(504,-60)(504,-54)
    \Line(492,-60)(492,-54)
    \Line(468,-60)(468,-54)
    \Line(456,-60)(456,-54)
    \Line(444,-60)(444,-54)
    \Line(432,-60)(432,-54)
    \Line(408,-60)(408,-54)
    \Text(540,-80)[lb]{\Large{\Black{$0$}}}
    \Text(600,-80)[lb]{\Large{\Black{$5$}}}
    \Text(650,-80)[lb]{\Large{\Black{$10$}}}
    \Text(378,18)[lb]{\Large{\Black{$0$}}}
    \Text(470,-80)[lb]{\Large{\Black{$-5$}}}
    \Text(400,-80)[lb]{\Large{\Black{$-10$}}}
    \Text(72,72)[lb]{\Large{\Black{$p_1$}}}
    \Text(78,-78)[lb]{\Large{\Black{$p_2$}}}
    \SetColor{WildStrawberry}
    \Line(185,61)(185,-31)
    \SetColor{Rhodamine}
    \Line[arrow,arrowpos=0.5,arrowlength=5,arrowwidth=2,arrowinset=0.2](180,59)(280,88)
    \SetColor{NavyBlue}
    \Line[arrow,arrowpos=0.5,arrowlength=5,arrowwidth=2,arrowinset=0.2](181,-34)(291,-66)
    \Text(193,13)[lb]{\Large{\Black{$\mathds{P}$}}}
    \Text(280,76)[lb]{\Large{\Black{${p_1}^{'}$}}}
    \SetColor{Rhodamine}
    \Vertex(406,62){3.606}
    \SetColor{NavyBlue}
    \Line[arrow,arrowpos=0.5,arrowlength=5,arrowwidth=2,arrowinset=0.2](180,-34)(285,-49)
    \Line[arrow,arrowpos=0.5,arrowlength=5,arrowwidth=2,arrowinset=0.2](180,-34)(260,-74)
    \Line[arrow,arrowpos=0.5,arrowlength=5,arrowwidth=2,arrowinset=0.2](180,-34)(280,-34)
    \Line[arrow,arrowpos=0.5,arrowlength=5,arrowwidth=2,arrowinset=0.2](180,-34)(285,-19)
    \SetColor{Black}
    \ZigZag(590,-59)(585,86){7.5}{7}
    \SetColor{NavyBlue}
    \Vertex(595,61){3.606}
    \Vertex(655,36){3.606}
    \Vertex(630,-29){3.606}
    \Vertex(650,71){3.606}
    \Vertex(625,41){3.606}
    \Vertex(650,1){3.606}
    \Vertex(605,-4){3.606}
	 \Text(325,-54)[lb]{\Large{\Black{$X_2$}}}
    \SetColor{Black}
	 \Bezier(298,-17)(314,-19)(314,-45)(320,-50)\Line(312.667,-32.154)(312.75,-32.375)
    \Bezier(300,-84)(312,-84)(313,-56)(319,-50)\Line(311.675,-69.482)(311.75,-69.25)
  \end{picture}
}
}

%% file: plots/DiffPom/proc_dd_fig.tex

\resizebox{4in}{!}{
\fcolorbox{white}{white}{
  \begin{picture}(666,191) (65,-75)
    \SetWidth{1.0}
    \SetColor{Black}
    \Line[arrow,arrowpos=0.5,arrowlength=5,arrowwidth=2,arrowinset=0.2](66,95)(180,63)
    \Line[arrow,arrowpos=0.5,arrowlength=5,arrowwidth=2,arrowinset=0.2](66,-61)(180,-29)
    \SetColor{WildStrawberry}
    \Line(179,64)(180,-29)
    \SetColor{Black}
    \GBox(396,-55)(684,95){0.882}
    \Text(372,95)[lb]{\Large{\Black{$\phi$}}}
    \Text(696,-67)[lb]{\Large{\Black{$\eta$}}}
    \Line(396,19)(402,19)
    \Line(540,-55)(540,-43)
    \Line(600,-55)(600,-43)
    \Line(660,-55)(660,-43)
    \Line(480,-55)(480,-43)
    \Line(420,-55)(420,-43)
    \Line(552,-55)(552,-49)
    \Line(564,-55)(564,-49)
    \Line(576,-55)(576,-49)
    \Line(588,-55)(588,-49)
    \Line(612,-55)(612,-49)
    \Line(624,-55)(624,-49)
    \Line(636,-55)(636,-49)
    \Line(648,-55)(648,-49)
    \Line(672,-55)(672,-49)
    \Line(528,-55)(528,-49)
    \Line(516,-55)(516,-49)
    \Line(504,-55)(504,-49)
    \Line(492,-55)(492,-49)
    \Line(468,-55)(468,-49)
    \Line(456,-55)(456,-49)
    \Line(444,-55)(444,-49)
    \Line(432,-55)(432,-49)
    \Line(408,-55)(408,-49)
    \Text(536,-79)[lb]{\Large{\Black{$0$}}}
    \Text(593,-80)[lb]{\Large{\Black{$5$}}}
    \Text(650,-78)[lb]{\Large{\Black{$10$}}}
    \Text(377,6)[lb]{\Large{\Black{$0$}}}
    \Text(460,-77)[lb]{\Large{\Black{$-5$}}}
    \Text(401,-78)[lb]{\Large{\Black{$-10$}}}
    \Text(72,77)[lb]{\Large{\Black{$p_1$}}}
    \Text(78,-73)[lb]{\Large{\Black{$p_2$}}}
    \SetColor{WildStrawberry}
    \Line(183,64)(183,-28)
    \SetColor{NavyBlue}
    \Line[arrow,arrowpos=0.5,arrowlength=5,arrowwidth=2,arrowinset=0.2](181,64)(281,93)
    \Line[arrow,arrowpos=0.5,arrowlength=5,arrowwidth=2,arrowinset=0.2](181,-29)(291,-61)
    \Text(193,18)[lb]{\Large{\Black{$\mathds{P}$}}}
    \SetWidth{0.0}
    \Vertex(420,58){4.243}
    \Vertex(639,17){4.243}
    \SetWidth{1.0}
    \Line[arrow,arrowpos=0.5,arrowlength=5,arrowwidth=2,arrowinset=0.2](179,-28)(282,-39)
    \Line[arrow,arrowpos=0.5,arrowlength=5,arrowwidth=2,arrowinset=0.2](179,-27)(264,-69)
    \Line[arrow,arrowpos=0.5,arrowlength=5,arrowwidth=2,arrowinset=0.2](179,-29)(276,-25)
    \Line[arrow,arrowpos=0.5,arrowlength=5,arrowwidth=2,arrowinset=0.2](183,63)(283,49)
    \Line[arrow,arrowpos=0.5,arrowlength=5,arrowwidth=2,arrowinset=0.2](185,63)(282,62)
    \Line[arrow,arrowpos=0.5,arrowlength=5,arrowwidth=2,arrowinset=0.2](185,65)(276,80)
    \Line[arrow,arrowpos=0.5,arrowlength=5,arrowwidth=2,arrowinset=0.2](180,63)(250,95)
    \SetWidth{0.0}
    \Vertex(419,-7){4.243}
    \Vertex(494,48){4.243}
    \Vertex(463,-24){4.243}
    \Vertex(451,29){4.243}
    \Vertex(443,69){4.243}
    \Vertex(589,-16){4.243}
    \Vertex(640,63){4.243}
    \Vertex(630,-24){4.243}
    \Vertex(609,33){4.243}
    \Vertex(665,8){4.243}
    \SetWidth{1.0}
    \SetColor{Black}
    \ZigZag(491,-48)(484,94){7.5}{7}
    \ZigZag(592,-54)(590,92){7.5}{7}
    \Text(310,78)[lb]{\Large{\Black{$X_1$}}}
    \Text(315,-54)[lb]{\Large{\Black{$X_2$}}}
    \Bezier(287,-17)(297,-17)(297,-40)(304,-40)\Line(296.561,-28.328)(296.625,-28.5)
    \Bezier(291,-74)(300,-73)(298,-48)(302,-42)\Line(298.341,-60.089)(298.375,-59.875)
    \Bezier(286,108)(295,109)(295,89)(300,87)\Line(294.447,98.779)(294.5,98.625)
    \Bezier(290,49)(299,49)(296,78)(300,87)\Line(296.849,64.374)(296.875,64.625)

	 \end{picture}
}
}

%% file: plots/DiffPom/proc_cd_fig.tex

\resizebox{4in}{!}{
\fcolorbox{white}{white}{
  \begin{picture}(666,191) (65,-75)
    \SetWidth{1.0}
    \SetColor{Black}
    \Line[arrow,arrowpos=0.5,arrowlength=5,arrowwidth=2,arrowinset=0.2](66,95)(180,63)
    \Line[arrow,arrowpos=0.5,arrowlength=5,arrowwidth=2,arrowinset=0.2](66,-61)(180,-29)
    \SetColor{WildStrawberry}
    \Line(179,64)(180,16)
    \SetColor{Black}
    \GBox(396,-55)(684,95){0.882}
    \Text(372,95)[lb]{\Large{\Black{$\phi$}}}
    \Text(696,-67)[lb]{\Large{\Black{$\eta$}}}
    \Line(396,19)(402,19)
    \Line(540,-55)(540,-43)
    \Line(600,-55)(600,-43)
    \Line(660,-55)(660,-43)
    \Line(480,-55)(480,-43)
    \Line(420,-55)(420,-43)
    \Line(552,-55)(552,-49)
    \Line(564,-55)(564,-49)
    \Line(576,-55)(576,-49)
    \Line(588,-55)(588,-49)
    \Line(612,-55)(612,-49)
    \Line(624,-55)(624,-49)
    \Line(636,-55)(636,-49)
    \Line(648,-55)(648,-49)
    \Line(672,-55)(672,-49)
    \Line(528,-55)(528,-49)
    \Line(516,-55)(516,-49)
    \Line(504,-55)(504,-49)
    \Line(492,-55)(492,-49)
    \Line(468,-55)(468,-49)
    \Line(456,-55)(456,-49)
    \Line(444,-55)(444,-49)
    \Line(432,-55)(432,-49)
    \Line(408,-55)(408,-49)
    \Text(536,-79)[lb]{\Large{\Black{$0$}}}
    \Text(593,-80)[lb]{\Large{\Black{$5$}}}
    \Text(650,-78)[lb]{\Large{\Black{$10$}}}
    \Text(377,6)[lb]{\Large{\Black{$0$}}}
    \Text(460,-77)[lb]{\Large{\Black{$-5$}}}
    \Text(401,-78)[lb]{\Large{\Black{$-10$}}}
    \Text(72,77)[lb]{\Large{\Black{$p_1$}}}
    \Text(78,-73)[lb]{\Large{\Black{$p_2$}}}
    \SetColor{WildStrawberry}
    \Line(183,64)(183,16)
    \SetColor{Rhodamine}
    \Line[arrow,arrowpos=0.5,arrowlength=5,arrowwidth=2,arrowinset=0.2](181,64)(281,93)
    \Line[arrow,arrowpos=0.5,arrowlength=5,arrowwidth=2,arrowinset=0.2](179,-30)(283,-48)
    \Text(192,39)[lb]{\Large{\Black{$\mathds{P}$}}}
    \SetWidth{0.0}
    \Vertex(414,60){4.243}
    \SetColor{NavyBlue}
    \Vertex(513,-28){4.243}
    \SetWidth{1.0}
    \Line[arrow,arrowpos=0.5,arrowlength=5,arrowwidth=2,arrowinset=0.2](184,15)(284,1)
    \Line[arrow,arrowpos=0.5,arrowlength=5,arrowwidth=2,arrowinset=0.2](184,16)(281,15)
    \Line[arrow,arrowpos=0.5,arrowlength=5,arrowwidth=2,arrowinset=0.2](184,16)(275,31)
    \Line[arrow,arrowpos=0.5,arrowlength=5,arrowwidth=2,arrowinset=0.2](185,16)(297,25)
    \SetWidth{0.0}
    \Vertex(494,48){4.243}
    \Vertex(537,10){4.243}
    \Vertex(589,-16){4.243}
    \Vertex(540,58){4.243}
    \Vertex(560,-21){4.243}
    \SetColor{Rhodamine}
    \Vertex(672,-17){4.243}
    \SetColor{NavyBlue}
    \Vertex(566,77){4.243}
    \SetWidth{1.0}
    \SetColor{Black}
    \ZigZag(498,-47)(491,95){7.5}{7}
    \ZigZag(579,-53)(577,93){7.5}{7}
    \Text(299,78)[lb]{\Large{\Black{${p_1}^{'}$}}}
    \Text(315,-54)[lb]{\Large{\Black{${p_2}^{'}$}}}
    \SetColor{WildStrawberry}
    \Line(184,16)(184,-32)
    \Line(180,18)(181,-30)
    \Text(192,-20)[lb]{\Large{\Black{$\mathds{P}$}}}
    \Text(316,19)[lb]{\Large{\Black{$X$}}}
    \SetColor{Black}
	 \Bezier(291,39)(303,39)(302,23)(309,22)\Line(301.811,30.999)(301.875,30.875)
    \Bezier(295,-2)(305,-1)(302,19)(310,21)\Line(303.205,8.964)(303.25,9.125)
  \end{picture}
}
}

%% file: plots/DiffPom/proc_nd_fig.tex

\resizebox{4in}{!}{
\fcolorbox{white}{white}{
  \begin{picture}(666,191) (65,-75)
    \SetWidth{1.0}
    \SetColor{Black}
    \Line[arrow,arrowpos=0.5,arrowlength=5,arrowwidth=2,arrowinset=0.2](66,95)(180,63)
    \Line[arrow,arrowpos=0.5,arrowlength=5,arrowwidth=2,arrowinset=0.2](66,-61)(180,-29)
    \SetColor{WildStrawberry}
    \Line(179,64)(179,-29)
    \SetColor{Black}
    \GBox(396,-55)(684,95){0.882}
    \Text(372,95)[lb]{\Large{\Black{$\phi$}}}
    \Text(696,-67)[lb]{\Large{\Black{$\eta$}}}
    \Line(396,19)(402,19)
    \Line(540,-55)(540,-43)
    \Line(600,-55)(600,-43)
    \Line(660,-55)(660,-43)
    \Line(480,-55)(480,-43)
    \Line(420,-55)(420,-43)
    \Line(552,-55)(552,-49)
    \Line(564,-55)(564,-49)
    \Line(576,-55)(576,-49)
    \Line(588,-55)(588,-49)
    \Line(612,-55)(612,-49)
    \Line(624,-55)(624,-49)
    \Line(636,-55)(636,-49)
    \Line(648,-55)(648,-49)
    \Line(672,-55)(672,-49)
    \Line(528,-55)(528,-49)
    \Line(516,-55)(516,-49)
    \Line(504,-55)(504,-49)
    \Line(492,-55)(492,-49)
    \Line(468,-55)(468,-49)
    \Line(456,-55)(456,-49)
    \Line(444,-55)(444,-49)
    \Line(432,-55)(432,-49)
    \Line(408,-55)(408,-49)
    \Text(536,-79)[lb]{\Large{\Black{$0$}}}
    \Text(593,-80)[lb]{\Large{\Black{$5$}}}
    \Text(650,-78)[lb]{\Large{\Black{$10$}}}
    \Text(377,6)[lb]{\Large{\Black{$0$}}}
    \Text(460,-77)[lb]{\Large{\Black{$-5$}}}
    \Text(401,-78)[lb]{\Large{\Black{$-10$}}}
    \Text(72,77)[lb]{\Large{\Black{$p_1$}}}
    \Text(78,-73)[lb]{\Large{\Black{$p_2$}}}
    \SetColor{NavyBlue}
    \Line[arrow,arrowpos=0.5,arrowlength=5,arrowwidth=2,arrowinset=0.2](181,64)(281,93)
    \Line[arrow,arrowpos=0.5,arrowlength=5,arrowwidth=2,arrowinset=0.2](179,-30)(283,-48)
    \SetWidth{0.0}
    \Vertex(414,60){4.243}
    \Vertex(513,-28){4.243}
    \SetWidth{1.0}
    \Line[arrow,arrowpos=0.5,arrowlength=5,arrowwidth=2,arrowinset=0.2](181,-8)(281,-22)
    \Line[arrow,arrowpos=0.5,arrowlength=5,arrowwidth=2,arrowinset=0.2](180,16)(277,15)
    \Line[arrow,arrowpos=0.5,arrowlength=5,arrowwidth=2,arrowinset=0.2](180,35)(271,50)
    \Line[arrow,arrowpos=0.5,arrowlength=5,arrowwidth=2,arrowinset=0.2](180,16)(292,25)
    \SetWidth{0.0}
    \Vertex(503,40){4.243}
    \Vertex(537,10){4.243}
    \Vertex(589,-16){4.243}
    \Vertex(540,58){4.243}
    \Vertex(560,-21){4.243}
    \Vertex(672,-17){4.243}
    \Vertex(566,77){4.243}
    \SetWidth{1.0}
    \Line[arrow,arrowpos=0.5,arrowlength=5,arrowwidth=2,arrowinset=0.2](178,64)(261,72)
    \Line[arrow,arrowpos=0.5,arrowlength=5,arrowwidth=2,arrowinset=0.2](176,35)(271,42)
    \Line[arrow,arrowpos=0.5,arrowlength=5,arrowwidth=2,arrowinset=0.2](182,-9)(251,-7)
    \Line[arrow,arrowpos=0.5,arrowlength=5,arrowwidth=2,arrowinset=0.2](178,-29)(233,-33)
    \Line[arrow,arrowpos=0.5,arrowlength=5,arrowwidth=2,arrowinset=0.2](181,-29)(231,-45)
    \Line[arrow,arrowpos=0.5,arrowlength=5,arrowwidth=2,arrowinset=0.2](181,35)(216,33)
    \Line[arrow,arrowpos=0.5,arrowlength=5,arrowwidth=2,arrowinset=0.2](181,14)(226,10)
    \SetWidth{0.0}
    \Vertex(654,65){4.243}
    \Vertex(466,68){4.243}
    \Vertex(616,26){4.243}
    \Vertex(600,63){4.243}
    \Vertex(473,19){4.243}
    \Vertex(435,-3){4.243}
  \end{picture}
}
}

%% file: plots/DiffPom/SD.tex
 \resizebox{4in}{!}{
  \fcolorbox{white}{white}{
  \begin{picture}(441,297) (137,-90)
    \SetWidth{2.0}
    \SetColor{Black}
    \Line[arrow,arrowpos=0.5,arrowlength=6.667,arrowwidth=2.667,arrowinset=0.2](161,174)(338,163)
    \Line[arrow,arrowpos=0.5,arrowlength=6.667,arrowwidth=2.667,arrowinset=0.2](338,163)(460,165)
    \SetWidth{1.3}
    \SetColor{Red}
    \ZigZag(333.934,164.813)(347.934,79.813){7.5}{4}\ZigZag(324.066,163.187)(338.066,78.187){7.5}{4}
    \SetWidth{1.0}
    \SetColor{Black}
    \GOval(347,65)(12,14)(0){0.882}
    \SetColor{Blue}
    \Gluon(359,69)(452,62){7.5}{7}
    \SetColor{Black}
    \Gluon(354,54)(370,17){7.5}{3}
    \SetColor{Blue}
    \Gluon(369,14)(445,20){7.5}{6}
    \SetColor{Black}
    \Gluon(370,16)(358,-33){7.5}{4}
    \Line[arrow,arrowpos=0.5,arrowlength=5,arrowwidth=2,arrowinset=0.2](285,-45)(358,-33)
    \GOval(263,-57)(23,25)(0){0.882}
    \SetWidth{2.0}
    \Line[arrow,arrowpos=0.5,arrowlength=6.667,arrowwidth=2.667,arrowinset=0.2](157,-54)(239,-55)
    \SetWidth{1.0}
    \SetColor{Blue}
    \Line[arrow,arrowpos=0.5,arrowlength=5,arrowwidth=2,arrowinset=0.2](289,-58)(457,-63)
    \Line[arrow,arrowpos=0.5,arrowlength=5,arrowwidth=2,arrowinset=0.2](286,-66)(455,-83)
    \Text(139,186)[lb]{\Large{\Black{$p_i$}}}
    \Text(134,-50)[lb]{\Large{\Black{$p_j$}}}
    \Text(474,172)[lb]{\Large{\Black{$p_{i
	 }^{'}$}}}
    \Text(331,31)[lb]{\Large{\Black{$x_g$}}}
    \Text(303,117)[lb]{\Large{\Black{$x_{\mathds{P}}$}}}
    \Text(459,135)[lb]{\Large{\Black{$LRG$}}}
    \SetColor{Black}
    \Bezier(479,67)(505,63)(502,3)(527,-4)\Line(503.206,33.116)(503.375,32.625)
    \Bezier(488,-89)(508,-88)(506,-19)(527,-4)\Line(506.986,-52.327)(507.125,-51.75)
    \Text(543,-3)[lb]{\Large{\Black{$X$}}}
    \SetColor{Blue}
    \Line[arrow,arrowpos=0.5,arrowlength=5,arrowwidth=2,arrowinset=0.2](361,-32)(448,-37)
    \SetColor{Black}
    \Vertex(367,18){3.606}
    \Vertex(358,-32){3.606}
    \Vertex(327,165){4}
  \end{picture}
}
}

%% file: DiffractionPythia.tex
\section{Diffraction in PYTHIA}

\subsection{PYTHIA 6}

\noindent The development of series 6 of PYTHIA written in Fortran 77 began in 1997. Although there was significant development from one version to the next, the description of diffraction remained the same (in the two versions 6.2 and 6.4). In this section a description of the diffractive processes in PYTHIA 6 is presented. Also presented is a comparison of diffractive kinematic distributions produced using PYTHIA and an alternative Monte Carlo generator called PHOJET \cite{phojet}. 

\subsubsection{Event Generation} \label{eventgeneration}

\noindent The total hadronic cross-section for $AB \rightarrow \textrm{anything}$, $\sigma_{\textrm{tot}}^{\textrm{AB}}$ is calculated using the Donnachie and Landshoff parameterization \cite{landshoff}. In this approach, the total cross-section appears as a sum of a Pomeron term and a Reggeon term, as seen in equation \ref{E:sigmatot}.

\begin{equation}\label{E:sigmatot}
  \sigma_{\textrm{tot}}^{\textrm{AB}}(s)=X^{\textrm{AB}}s^{\epsilon}+Y^{\textrm{AB}}s^{-\eta}
\end{equation}
\\
\noindent The powers $\epsilon$ for the Pomeron term and $\eta$ for the Reggeon term are expected to be universal, while the coefficients $X$ and $Y$ are specific to each initial state. Those parameterizations not provided in \cite{landshoff} have been calculated in \cite{schulersjostrand}. Cross-sections for elastic, single and double diffractive events are provided, but higher diffractive topologies like central diffraction are neglected. The diffractive cross-sections and event characteristics are described by a model by Schuler and Sj\"{o}strand found in \cite{schulersjostrand, schulersjostrand2}. The non diffractive cross section is given by ``whatever is left".\\

\noindent In the Schuler-Sj\"{o}strand model, when the square of the momentum transfer $t$ is not too large, the differential elastic cross section can be approximated by a simple exponential fall-off with respect to \textit{t}.  
\noindent Diffractive cross-sections have an inverse diffractive mass squared ($M^2$) dependence and an exponential dependence on $t$. The simple $dM^2/M^2$ form is modified by the mass dependence of the slope of the $t$ distribution (with co-efficients $B_{\textrm{sd(XB)}}$, $B_{\textrm{sd(AX)}}$ and $B_{\textrm{dd}}$) \cite{pythia6.4}. These Regge formulae for diffraction are supposed to hold in certain asymptotic regions of the full phase space. For example, in $\textrm{p}+\textrm{p}\rightarrow \textrm{p}+\textrm{M}$, $|t|_{\textrm{min}}^{1/2}\approx m_{\textrm{p}}(M^2-m_{\textrm{p}}^2)/s$ \cite{schulersjostrand2}. Having a lower cut on \textit{t} of the order of $m_{\pi}^2$ implies $M^2-m_{\textrm{p}}^2<0.15s$. But there will be diffraction even outside these regions where the Regge formulae were derived. Due to the lack of a theory that predicts differential cross-sections at arbitrary $t$ and $M^2$ values, the above Regge formulae are used everywhere along with ``fudge factors" $F_{\textrm{sd}}$ and $F_{\textrm{dd}}$ in equations \ref{E:sdcs} and \ref{E:ddcs}. The form of these factors is given by equations 7.76 of \cite{pythia6.4} to give sensible behaviour in full phase space. These factors suppress production close to the kinematical limit and in the case of double diffraction, also suppresses configurations where the two diffractive systems overlap in rapidity space. These ``fudge factors" also give a broad enhancement in the production rate in the resonance region up to about \unit[2]{GeV}. This gives a smeared-out version of exclusive states, rather than listing them all out individually. 

\noindent Diffractive cross-sections are given by equations \ref{E:sdcs} and \ref{E:ddcs}. 

\begin{equation}\label{E:sdcs}
  \frac{d\sigma_{\textrm{sd}(\textrm{AB}\rightarrow \textrm{XB})}(s)}{dtdM^2}=\frac{g_{3\mathds{P}}}{16\pi}\beta_{\textrm{A}\mathds{P}}\beta^2_{\textrm{B}\mathds{P}}\frac{1}{M^2}\exp(B_{\textrm{sd(XB)}}t)F_{\textrm{sd}}
\end{equation}

\begin{equation}\label{E:ddcs}
  \frac{d\sigma_{\textrm{dd}}(s)}{dtdM_1^2dM_2^2}=\frac{g^2_{3\mathds{P}}}{16\pi}\beta_{\textrm{A}\mathds{P}}\beta_{\textrm{B}\mathds{P}}\frac{1}{M_1^2}\frac{1}{M_2^2}\exp(B_{\textrm{dd}}t)F_{\textrm{dd}}
\end{equation}

\vspace{10mm}

\noindent The couplings $\beta$ are related to the Pomeron term of equation \ref{E:sigmatot}. The triple Pomeron coupling $g_{3\mathds{P}}$ is determined from single diffractive data. The diffractive mass spectrum $M$ ranges from \unit[0.28]{GeV} ($\approx 2m_{\pi}$) above the mass of the diffracted hadron, to the kinematic limit. The exponential slope parameters $B_{\textrm{sd or dd}}$ are assumed to have a logarithmic dependence on $1/M^2$. The kinematic range of $t$ depends on the masses of all incoming and outgoing systems involved. More information and the equations can be found in section 7.7.1 of the PYTHIA 6.4 manual \cite{pythia6.4}.\\

\noindent Diffractive cross-sections have been integrated for a set of centre of mass (CM) energies, starting at \unit[10]{GeV}. The results have been parameterized in section 4 of \cite{schulersjostrand2}. Once the process is selected using this parameterization, $M$ and $t$ are generated using equations \ref{E:sdcs} and \ref{E:ddcs}. \\

\subsubsection{Particle Production}\label{particleproduction}

\noindent Once the process is selected and the kinematic variables are determined, the products of the collision are generated. The handling of this production depends on the relative value of the diffractive mass $M$. If $M\leq \unit[1]{GeV}$ above the mass of the incoming particle, the system is allowed to decay isotropically to a two-body system. For a more massive diffractive state, the system is treated as a string with the quantum numbers of the original hadron. Two alternative ways of stretching the string are considered.\\

\begin{figure}[htbp]
  \begin{center}
    \begin{tabular}{cc}
		\input{plots/DiffPyt/string_quark} & \input{plots/DiffPyt/string_gluon}\\
      (a) & (b)\\
    \end{tabular}
    \caption{\footnotesize{String being stretched in diffractive processes - (a) $\mathds{P}$ couples to a valence quark and (b) $\mathds{P}$ couples to a gluon}}
    \label{fig:string}
  \end{center}
\end{figure}
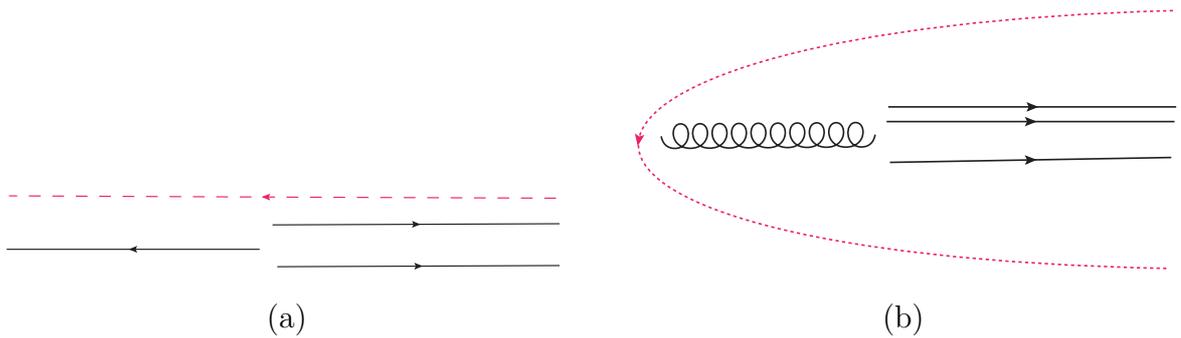

\noindent There is both a gluonic and a quark contribution. When an incoming hadron is diffractively excited, either a valence quark or a gluon is ``kicked out" of it. If the Pomeron couples to a valence quark of the non-diffracted proton, the string (the pink dashed lines in figure \ref{fig:string}) is stretched between the struck quark and the remnant diquark (or antiquark) of the diffractive system, seen in figure \ref{fig:string}(a). This configuration dominates at small $M$. The alternative is when the interaction is with a gluon of the non-diffracted proton. The string is stretched from a quark in the diffractive state to a gluon, and then back to a diquark (or antiquark). This gives rise to a hair-pin structure as seen in figure \ref{fig:string}(b). In PYTHIA 6 the ratio of the two contributions can be set. \\

\subsubsection{PYTHIA 6.214 vs PHOJET 1.12}

\noindent A study comparing the pseudorapidity ($\eta$), charged particle density ($dN_{\textrm{ch}}/d\eta$) and transverse momentum ($p_T$) distributions in PYTHIA 6.214 and PHOJET 1.12 at CM energy \unit[7]{TeV} is shown below. ND and SD spectra are compared to analyse the difference in the diffractive part. 

\begin{figure}[htbp]
        \begin{center}
		     \begin{tabular}{cc}
                (a) & \input{plots/DiffPyt/plot_etaND-6.tex} \\
                (b) & \input{plots/DiffPyt/plot_etaSD-6.tex} \\
			  \end{tabular}
                \caption{\footnotesize{$\eta$ distributions for (a) ND and (b) SD events at \unit[7]{TeV} comparing PYTHIA6 and PHOJET.}}
                \label{fig:eta_old}
        \end{center}
\end{figure}
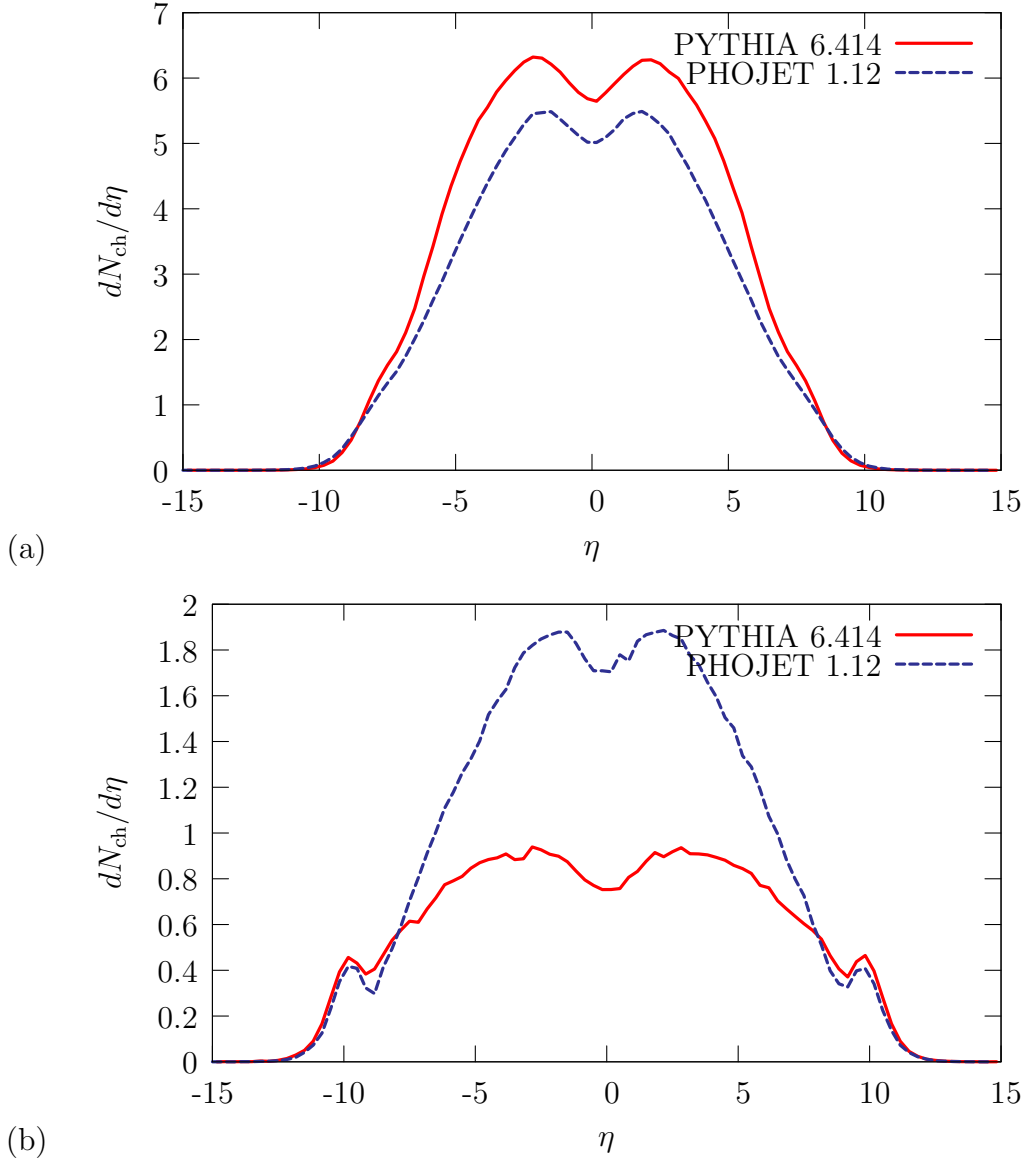

\begin{figure}[htbp]
        \begin{center}
		     \begin{tabular}{cc}
                (a) & \input{plots/DiffPyt/plot_multND-6.tex} \\
                (b) & \input{plots/DiffPyt/plot_multSD-6.tex} \\
            \end{tabular}
                \caption{\footnotesize{Multiplicity distributions for (a) ND and (b) SD events at \unit[7]{TeV} comparing PYTHIA6 and PHOJET.}}
                \label{fig:mult_old}
        \end{center}
\end{figure}

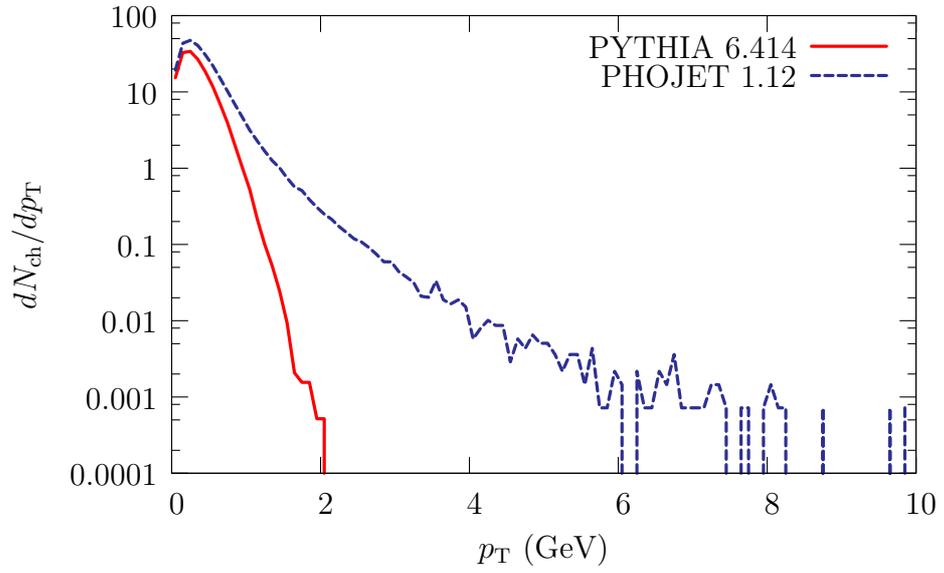
\begin{figure}[htbp]
        \begin{center}
		     \begin{tabular}{cc}
                (a) & \input{plots/DiffPyt/plot_pTND-6.tex} \\
                (b) & \input{plots/DiffPyt/plot_pTSD-6.tex} \\
			  \end{tabular}
                \caption{\footnotesize{$p_T$ distributions for (a) ND and (b) SD events at \unit[7]{TeV} comparing PYTHIA6 and PHOJET.}}
                \label{fig:pt_old}
        \end{center}
\end{figure}

\noindent A comparison of figures \ref{fig:eta_old}(a) with \ref{fig:eta_old}(b) and \ref{fig:mult_old}(a) with \ref{fig:mult_old}(b) shows that although the multiplicity spectra for ND events in PYTHIA and PHOJET are similar, high multiplicity SD events are not generated by PYTHIA. Similarly, the $p_T$ spectra in figures \ref{fig:pt_old}(a) and \ref{fig:pt_old}(b) show that PYTHIA lacks a hard diffractive part.   

\subsection{PYTHIA 8 before 8.130}\label{PYTHIA8}

\noindent PYTHIA 6.4 was the last version of PYTHIA to be coded in Fortran 77, followed by a switch to C++ with version 8.1. The mechanism for diffractive scattering works in almost the same way as in PYTHIA 6. The only difference lies in the particle production. In PYTHIA 8.1 the ratio of the probability of the Pomeron coupling to a quark ($P(\textrm{q})$) and the Pomeron coupling to a gluon ($P(\textrm{g})$) is given by equation \ref{E:string8}. $N$ in this equation is a normalization factor and $p$ (default value = 1) is a user-defined power. This introduces a mass dependence on the ratio of the two couplings, enabling the gluonic contribution to dominate at higher diffractive masses.

\begin{equation}\label{E:string8}
  \frac{P(\textrm{q})}{P(\textrm{g})}=\frac{N}{M^{p}}
\end{equation}

\subsection{PYTHIA 8.130}

\noindent In the versions of PYTHIA following PYTHIA 8.130, diffraction is
modelled based on the Pomeron approach described in section 1.2
. Pomeron-proton collisions are modeled at a reduced CM energy, then fully integrated
in such a way that it uses the standard PYTHIA machinery for multiple
interactions, parton showers and hadronization. This is the approach pioneered
in the POMPYT program \cite{pompyt}.\\

\subsubsection{Event Generation}
	
\noindent Diffractive cross sections are determined in exactly the same way as described in section \ref{eventgeneration}. However, in addition to the Schuler-Sj\"{o}strand model for picking $M$ and $t$, three other parameterizations of the Pomeron flux have been implemented. \\

\noindent 1. Bruni and Ingelman \cite{ingelmanbruni}: based on a critical Pomeron giving a mass spectrum close to $dM^2/M^2$. The $t$ distribution is the sum of two exponentials.\\


\noindent 2. Berger \textit{et al.} \cite{berger} and Streng \cite{streng}: a conventional Pomeron description but with values (from the RAPGAP manual \cite{rapgap}) updated to a super-critical Pomeron. This gives a stronger peaking towards low-mass diffractive states. The $t$ slope is still exponential and depends on $M$.\\ 


\noindent 3. Donnachie and Landshoff \cite{donlandshoff}: similar to the Streng-Berger parameterization, but with a power law distribution for $t$.\\ 

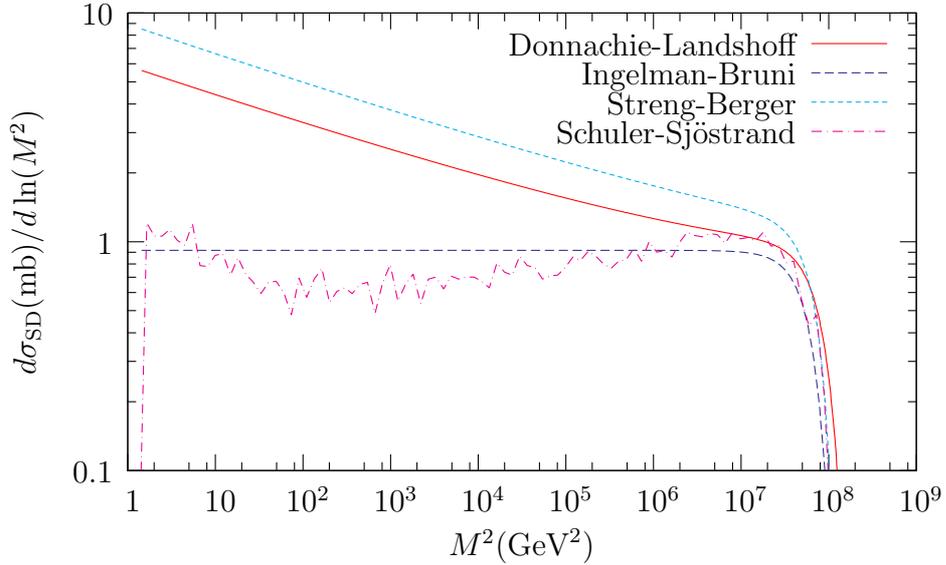
\begin{figure}[htbp]
        \begin{center}
					 \input{plots/DiffPyt/plot_lnM2.tex}
                \caption{\footnotesize{Differential cross-sections using different Pomeron fluxes.}}
                \label{fig:Pflux_x}
        \end{center}
\end{figure}

\noindent A comparison of the 4 different Pomeron fluxes are seen in figure \ref{fig:Pflux_x}. 

\noindent The Schuler-Sj\"{o}strand model is currently the only one which provides a separate $t$ spectrum for DD. 

\subsubsection{Particle Production}

\noindent PYTHIA by default only allows collisions with CM energy above \unit[10]{GeV}. But the diffractive mass spectra extend down to about \unit[1.2]{GeV}. A perturbative description at this scale is not possible, giving rise to a separate handling of low and high masses. For $M\leq\unit[10]{GeV}$, the non-perturbative description with longitudinally stretched strings, as described in sections \ref{particleproduction} and \ref{PYTHIA8} is implemented.  \\

\paragraph{High-Mass Diffraction}

\noindent In the mass range $\unit[10]{GeV}<M<\sqrt{s}$, a perturbative description is implemented. The probability for this description is given by 
$$ P_{\textrm{pert}}=1-\exp((m_{\textrm{diff}}-m_{\textrm{min}})/m_{\textrm{width}})$$
where $m_{\textrm{min}}$ and $m_{\textrm{width}}$ are free parameters. The default value of $m_{\textrm{min}}$ is set at \unit[10]{GeV} so that $P_{\textrm{pert}}$ vanishes when $M<\unit[10]{GeV}$.  

\noindent The perturbative description involves using PDFs for the Pomeron that are not well known. PYTHIA 8.130 provides a selection of five PDF sets. \\ 

\noindent 1. $Q^2$-independent parameterizations of the form given by equation \ref{qindepPDF}. Here $N$ is a normalization factor that ensures unit momentum sum and $a$ and $b$ can be different for the quark and gluonic content of the Pomeron. In this PDF set, the momentum fraction of gluons and quarks can be freely mixed. Additionally, the production of $s$ quarks can be suppressed relative to $u$ and $d$ quarks, with quarks and anti-quarks being equally likely to be produced.

\begin{equation}\label{qindepPDF}
  xf(x)=N_{\textrm{ab}}x^a(1-x)^b 
\end{equation}
\\
\noindent 2. A Pomeron can be described by the PDF for a pion. A few PDF sets exist, one of which is built into PYTHIA. The others can be accessed from LHAPDF \cite{LHAPDF}. Parameterizations are given for $\pi^+$; $\pi^-$ is obtained by charge conjugation and $\pi^0$ by averaging. \\
		
\noindent 3. The H1 2006 Fit A parameterization is a $Q^2$-dependent set. This is based on a tune (tune A) to H1 data on inclusive diffractive cross-section, described in section 5.3 of \cite{H12006}.\\
		
\noindent 4. The H1 2006 Fit B parameterization is another $Q^2$-dependent set based on tune B to the H1 data on inclusive diffractive cross-section, described in section 5.3 of \cite{H12006}.\\
		
\noindent 5. The H1 2007 Jets parameterization is a $Q^2$-dependent set based on a tune to H1 data. This fit uses measurements of both the difractive dijet cross-section presented in \cite{H12007} and the inclusive diffractive cross-section presented in \cite{H12006}.\\

\noindent PDF sets 3,4 and 5 above are next to leading order (NLO) sets, which may make them less suited for MC applications. A leading order (LO) gluon might be more stable at small \textit{z} when evolving to lower scales and a LO set will attach better to the LO matrix elements of PYTHIA. A LO fit from the H1 collaboration \cite{H1LO} is due to be added to the list of PDF sets.\\

\noindent Parton distributions, by default, are normalized so that they obey the momentum sum rule 
$$ \int_0^1 \!zf(z) \,dz = 1.$$ The motivation for this to hold is described in \cite{momsum}. However, since the Pomeron is not a physical particle, DPDFs do not implement momentum sum rules. Those from H1 add up to a momentum sum of roughly 50\%. Only the product of the Pomeron flux and the Pomeron PDF is meaningful, allowing arbitary separate normalizations of the Pomeron flux and the Pomeron PDF. H1 choose to normalize their flux so that 
$$ f_{\mathds{P}/p}(x_{\mathds{P}},t)=1 \ \ \textrm{when} \ \ x_{\mathds{P}}=0.003.$$

\noindent The standard perturbative multiple interactions framework for pp collisions provides parton-parton interaction cross-sections at a fixed CM energy. To turn these cross-sections into probabilities, one needs an ansatz for the Pomeron-proton total cross section. The single diffractive cross-section is given by equation \ref{E:SDcs}.

\begin{equation}\label{E:SDcs}
  \sigma_{\textrm{SD}}=\int\int dx_{\mathds{P}} dt \ f_{\mathds{P}/p}(x_{\mathds{P}},t) \ \ \underbrace{\sigma_{\mathds{P}p} (M^2=x_{\mathds{P}}s)\  \textrm{ . gap survival}}_{\sigma_{\mathds{P}p}(\textrm{effective})}
\end{equation}
		\\
\noindent The normalization of the Pomeron flux ($f_{\mathds{P}/p}(x_{\mathds{P}},t)$) is arbitrary and $\sigma_{\textrm{SD}}$ is parameterized from Regge theory. Then $\sigma_{\mathds{P}p}(\textrm{effective})$ is adjusted accordingly. The value of $\sigma_{\mathds{P}p}$ often quoted in literature is around \unit[2]{mb} \cite{berger}. In PYTHIA the default value of $\sigma_{\mathds{P}\textrm{p}}(\textrm{effective})$ is \unit[10]{mb}, which takes into account screening effects. This value is also used for multiple interactions in diffractive systems as described below. It is the main free tunable parameter in high-mass diffraction, and along with the choice of Pomeron PDF, can be fitted to represent diffractive event-shape data such as average charged multiplicity. The gap survival probability depends on the energy of the collision. The higher the energy, the greater the probability of multiple interactions in the same event that suppress the rapidity gap.\\
	
\noindent To describe the dampening of the perturbative jet cross-section at $p_T\rightarrow0$ by colour screening, the actual cross-section is multiplied by a regularization factor $p_T^4/(p_{T0}^2+p_T^2)^2$. $p_{T0}$ is a free, tunable parameter of the order \unit[2-4]{GeV}. The energy dependence of $p_{T0}$ is given by 
$$p_{T0}(\textrm{ecmNow})=p_T(\textrm{Ref})\left(\frac{\textrm{ecmNow}}{\textrm{ecmRef}}\right)^{\textrm{ecmPow}}$$
\noindent where ``ecmNow" is the current energy scale, ``ecmRef" is an arbitrary reference energy at which $p_T(\textrm{Ref})=p_{T0}(\textrm{ecmRef})$ is defined and ``ecmPow" is the energy rescaling pace. \\

\noindent Integrating equation \ref{E:pPcs} gives the total minijet pair cross-section. The average number of jet pairs in an event is given by $\frac{\textrm{total minijet pair cross-section}}{\sigma_{\mathds{P}\textrm{p}}(\textrm{effective})}$. Therefore, increasing the value of $\sigma_{\mathds{P}\textrm{p}}(\textrm{effective})$ will reduce the multiple interactions activity per event. This also explains the choice of $\sigma_{\mathds{P}p}(\textrm{effective})$ above. \\

\noindent Even at a fixed CM energy, the diffractive (high) mass spectrum can lie in the range $\unit[10]{GeV}<M<\sqrt{s}$, with a varying set of parameters (such as the $p_T$ cut-off parameter ($p_{T0}$)) along the range. Therefore, multiple interactions are initialised for a few (currently five) different diffractive mass values across the range, and all relevant parameters are interpolated between them to obtain the behaviour at a specific diffractive mass. Additionally, $AB\rightarrow XB$ and $AB\rightarrow AX$ are initialized separately. This allows for different beams (or PDFs) on both sides. This also facilitates double diffraction.  \\

\subsection{PYTHIA 8.130 vs PYTHIA 6.214 and PHOJET 1.12}

\noindent A study comparing the pseudorapidity ($\eta$), transverse momentum ($p_T$) and charged particle density ($dN_{\textrm{ch}}/d\eta$) distributions in PYTHIA 8.130, PYTHIA 6.214 and PHOJET 1.12 at CM energy \unit[7]{TeV} is shown below. Only the SD spectra are compared. 

\begin{figure}[htbp]
        \begin{center}
               \input{plots/DiffPyt/plot_etaSD.tex}
                \caption{\footnotesize{$\eta$ distribution for SD events at \unit[7]{TeV} using PYTHIA8.}}
                \label{fig:eta_new}
        \end{center}
\end{figure}
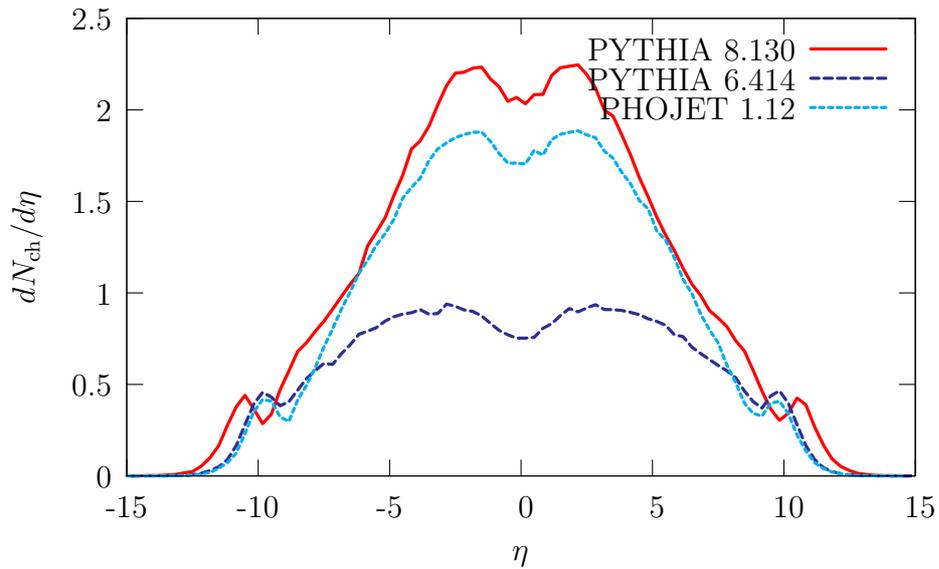

\begin{figure}[htbp]
        \begin{center}
                \input{plots/DiffPyt/plot_pTSD.tex}
                \caption{\footnotesize{$p_T$ distribution for SD events at \unit[7]{TeV} using PYTHIA8.}}
                \label{fig:pt_new}
        \end{center}
\end{figure}
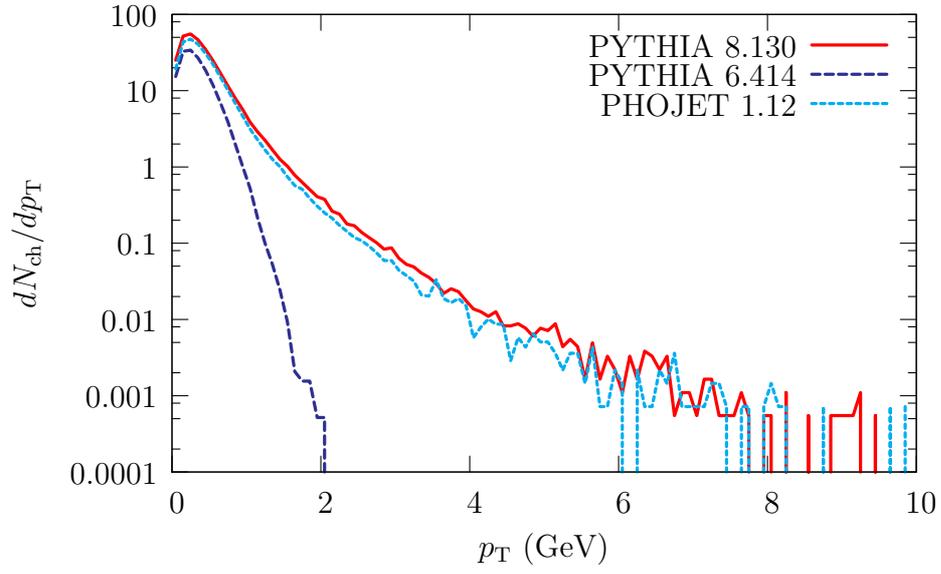

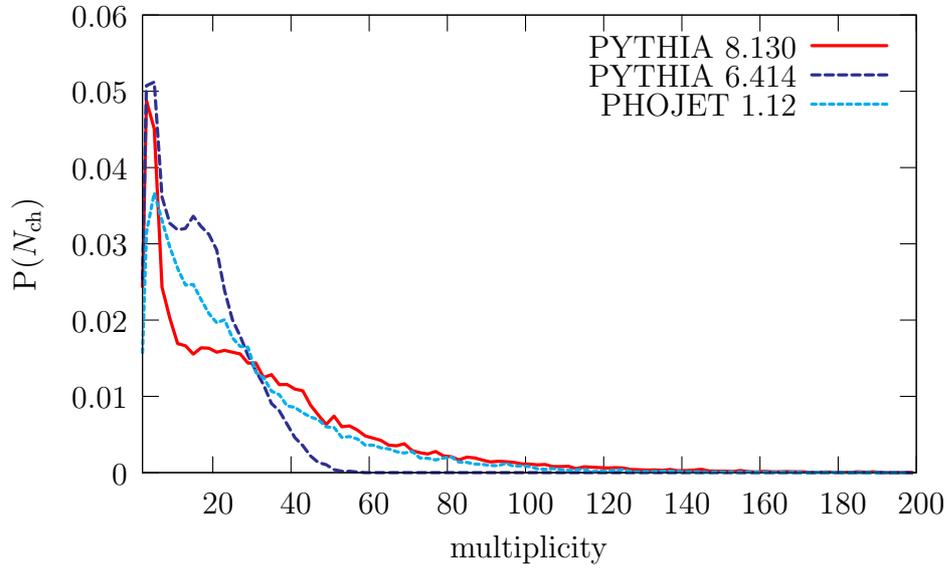
\begin{figure}[htbp]
        \begin{center}
                \input{plots/DiffPyt/plot_multSD.tex}
                \caption{\footnotesize{multiplicity distribution for SD events at \unit[7]{TeV} using PYTHIA8.}}
                \label{fig:mult_new}
        \end{center}
\end{figure}

\noindent It is clearly seen that the addition of hard diffraction to PYTHIA shows an improvement in the $p_T$ and multiplicity tails, giving a description comparable to PHOJET, which also has hard diffractive scattering.

%% file: plots/DiffPyt/string_quark.tex
\resizebox{3in}{!}{
	\fcolorbox{white}{white}{
		\begin{picture}(474,66) (120,-122)
      	\SetWidth{1.0}
         \SetColor{Black}
			\Line[arrow,arrowpos=0.5,arrowlength=5,arrowwidth=2,arrowinset=0.2](348,-83)(593,-82)
         \Line[arrow,arrowpos=0.5,arrowlength=5,arrowwidth=2,arrowinset=0.2](352,-119)(593,-118)
         \Line[arrow,arrowpos=0.5,arrowlength=5,arrowwidth=2,arrowinset=0.2](337,-104)(121,-104)
         \SetColor{WildStrawberry}
			\DashBezier(590,-60)(372,-59)(301,-59)(123,-58){10}\Line[arrow,arrowpos=0.5,arrowlength=5,arrowwidth=2,arrowinset=0.2](343.519,-59.008)(341.5,-59)
     	\end{picture}
	}  
}  

%% file: plots/DiffPyt/string_gluon.tex
\resizebox{3in}{!}{
	\fcolorbox{white}{white}{
   	\begin{picture}(331,160) (189,-69)
      	\SetWidth{1.0}
         \SetColor{Black}
         \Line[arrow,arrowpos=0.5,arrowlength=5,arrowwidth=2,arrowinset=0.2](343,22)(518,22)
         \Line[arrow,arrowpos=0.5,arrowlength=5,arrowwidth=2,arrowinset=0.2](345,-3)(516,0)
         \Line[arrow,arrowpos=0.5,arrowlength=5,arrowwidth=2,arrowinset=0.2](344,31)(519,31)
         \Gluon(336,14)(206,13){7.5}{10}
         \SetColor{WildStrawberry}
         \DashBezier(517,90)(102,79)(67,-59)(514,-68){2}\Line[arrow,arrowpos=0.5,arrowlength=5,arrowwidth=2,arrowinset=0.2](192.425,11.36)(192.25,10.25)
   	\end{picture}
  	}
}

%% file: plots/DiffPyt/plot_etaND-6.tex
\begin{picture}(0,0)%
	\includegraphics{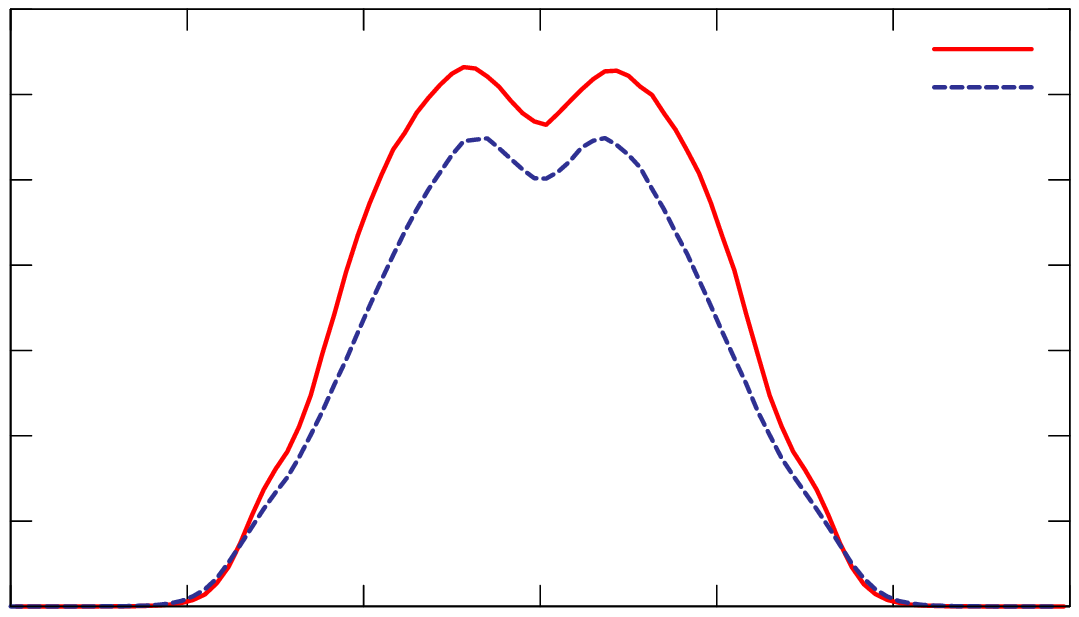}%
\end{picture}%
\begingroup
	\setlength{\unitlength}{0.0200bp}%
	\begin{picture}(18000,10800)(0,0)%
		\put(1650,1650){\makebox(0,0)[r]{\strut{} 0}}%
		\put(1650,2879){\makebox(0,0)[r]{\strut{} 1}}%
		\put(1650,4107){\makebox(0,0)[r]{\strut{} 2}}%
		\put(1650,5336){\makebox(0,0)[r]{\strut{} 3}}%
		\put(1650,6564){\makebox(0,0)[r]{\strut{} 4}}%
		\put(1650,7793){\makebox(0,0)[r]{\strut{} 5}}%
		\put(1650,9021){\makebox(0,0)[r]{\strut{} 6}}%
		\put(1650,10250){\makebox(0,0)[r]{\strut{} 7}}%
		\put(1925,1100){\makebox(0,0){\strut{}-15}}%
		\put(4467,1100){\makebox(0,0){\strut{}-10}}%
		\put(7008,1100){\makebox(0,0){\strut{}-5}}%
		\put(9550,1100){\makebox(0,0){\strut{} 0}}%
		\put(12092,1100){\makebox(0,0){\strut{} 5}}%
		\put(14633,1100){\makebox(0,0){\strut{} 10}}%
		\put(17175,1100){\makebox(0,0){\strut{} 15}}%
		\put(550,5950){\rotatebox{90}{\makebox(0,0){\strut{}$dN_{\mathrm{ch}}/d\eta$}}}%
		\put(9550,275){\makebox(0,0){\strut{}$\eta$}}%
		\put(14950,9675){\makebox(0,0)[r]{\strut{}PYTHIA 6.414}}%
		\put(14950,9125){\makebox(0,0)[r]{\strut{}PHOJET 1.12}}%
	\end{picture}%
\endgroup
 

%% file: plots/DiffPyt/plot_etaSD-6.tex
\begin{picture}(0,0)%
	\includegraphics{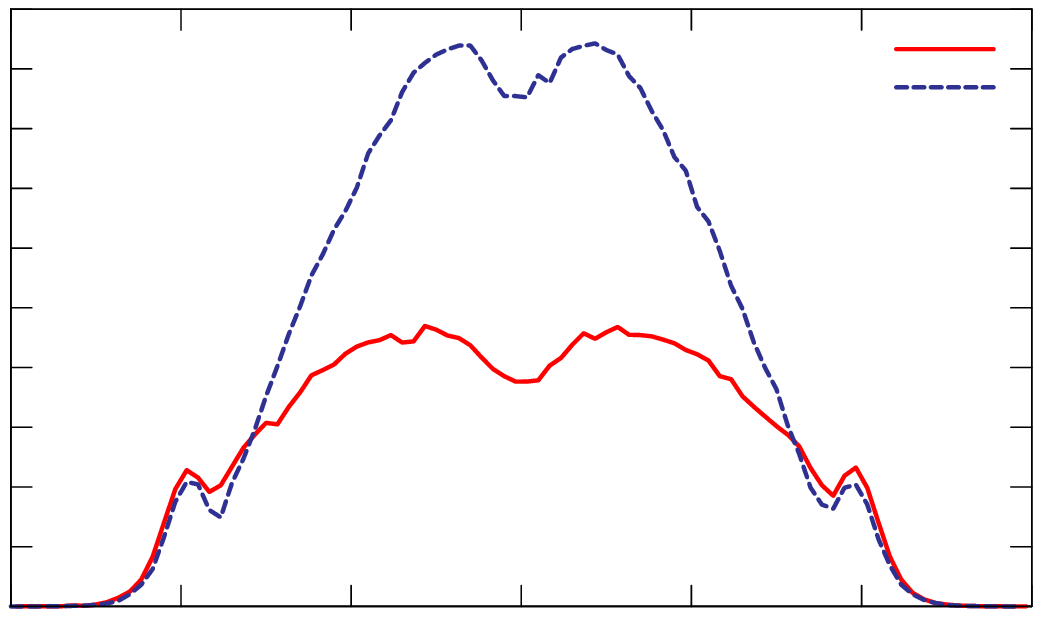}%
\end{picture}%
\begingroup
	\setlength{\unitlength}{0.0200bp}%
	\begin{picture}(18000,10800)(0,0)%
		\put(2200,1650){\makebox(0,0)[r]{\strut{} 0}}%
		\put(2200,2510){\makebox(0,0)[r]{\strut{} 0.2}}%
		\put(2200,3370){\makebox(0,0)[r]{\strut{} 0.4}}%
		\put(2200,4230){\makebox(0,0)[r]{\strut{} 0.6}}%
		\put(2200,5090){\makebox(0,0)[r]{\strut{} 0.8}}%
		\put(2200,5950){\makebox(0,0)[r]{\strut{} 1}}%
		\put(2200,6810){\makebox(0,0)[r]{\strut{} 1.2}}%
		\put(2200,7670){\makebox(0,0)[r]{\strut{} 1.4}}%
		\put(2200,8530){\makebox(0,0)[r]{\strut{} 1.6}}%
		\put(2200,9390){\makebox(0,0)[r]{\strut{} 1.8}}%
		\put(2200,10250){\makebox(0,0)[r]{\strut{} 2}}%
		\put(2475,1100){\makebox(0,0){\strut{}-15}}%
		\put(4925,1100){\makebox(0,0){\strut{}-10}}%
		\put(7375,1100){\makebox(0,0){\strut{}-5}}%
		\put(9825,1100){\makebox(0,0){\strut{} 0}}%
		\put(12275,1100){\makebox(0,0){\strut{} 5}}%
		\put(14725,1100){\makebox(0,0){\strut{} 10}}%
		\put(17175,1100){\makebox(0,0){\strut{} 15}}%
		\put(550,5950){\rotatebox{90}{\makebox(0,0){\strut{}$dN_{\mathrm{ch}}/d\eta$}}}%
		\put(9825,275){\makebox(0,0){\strut{}$\eta$}}%
		\put(14950,9675){\makebox(0,0)[r]{\strut{}PYTHIA 6.414}}%
		\put(14950,9125){\makebox(0,0)[r]{\strut{}PHOJET 1.12}}%
	\end{picture}%
\endgroup
 

%% file: plots/DiffPyt/plot_multND-6.tex
\begin{picture}(0,0)%
	\includegraphics{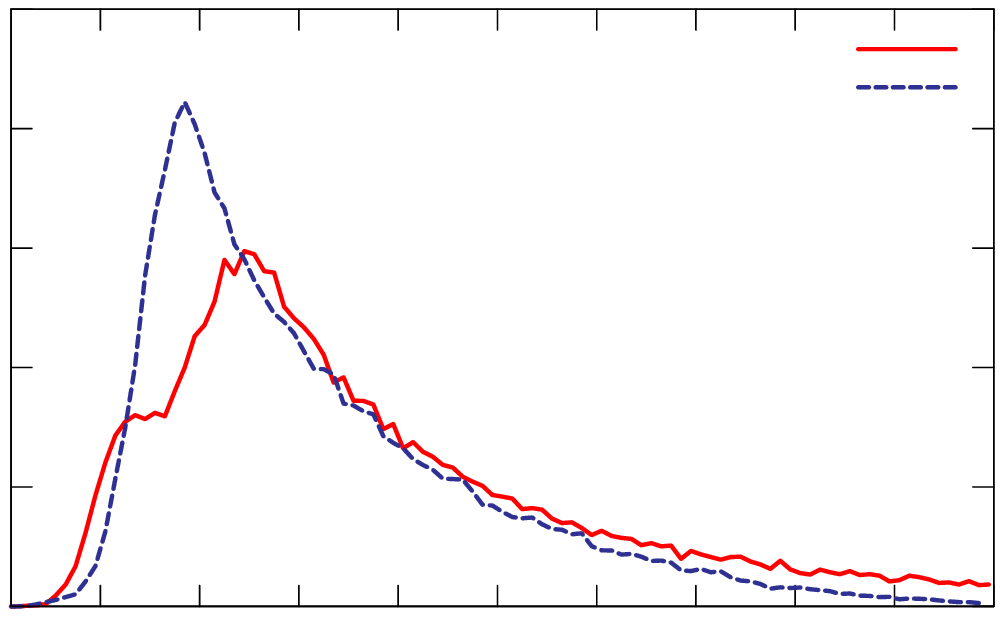}%
\end{picture}%
\begingroup
	\setlength{\unitlength}{0.0200bp}%
	\begin{picture}(18000,10800)(0,0)%
		\put(2750,1650){\makebox(0,0)[r]{\strut{} 0}}%
		\put(2750,3370){\makebox(0,0)[r]{\strut{} 0.005}}%
		\put(2750,5090){\makebox(0,0)[r]{\strut{} 0.01}}%
		\put(2750,6810){\makebox(0,0)[r]{\strut{} 0.015}}%
		\put(2750,8530){\makebox(0,0)[r]{\strut{} 0.02}}%
		\put(2750,10250){\makebox(0,0)[r]{\strut{} 0.025}}%
		\put(4311,1100){\makebox(0,0){\strut{} 20}}%
		\put(5741,1100){\makebox(0,0){\strut{} 40}}%
		\put(7170,1100){\makebox(0,0){\strut{} 60}}%
		\put(8599,1100){\makebox(0,0){\strut{} 80}}%
		\put(10029,1100){\makebox(0,0){\strut{} 100}}%
		\put(11458,1100){\makebox(0,0){\strut{} 120}}%
		\put(12887,1100){\makebox(0,0){\strut{} 140}}%
		\put(14316,1100){\makebox(0,0){\strut{} 160}}%
		\put(15746,1100){\makebox(0,0){\strut{} 180}}%
		\put(17175,1100){\makebox(0,0){\strut{} 200}}%
		\put(550,5950){\rotatebox{90}{\makebox(0,0){\strut{}$\mathrm{P}(N_{\mathrm{ch}})$}}}%
		\put(10100,275){\makebox(0,0){\strut{}multiplicity}}%
		\put(14950,9675){\makebox(0,0)[r]{\strut{}PYTHIA 6.414}}%
		\put(14950,9125){\makebox(0,0)[r]{\strut{}PHOJET 1.12}}%
	\end{picture}%
\endgroup
 

%% file: plots/DiffPyt/plot_multSD-6.tex
\begin{picture}(0,0)%
	\includegraphics{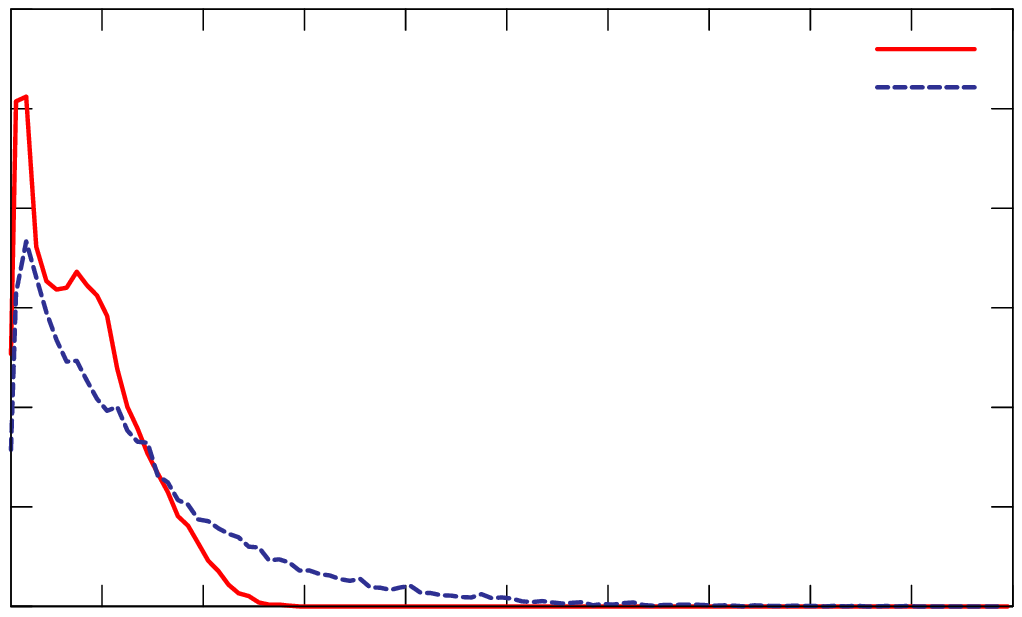}%
\end{picture}%
\begingroup
	\setlength{\unitlength}{0.0200bp}%
	\begin{picture}(18000,10800)(0,0)%
		\put(2475,1650){\makebox(0,0)[r]{\strut{} 0}}%
		\put(2475,3083){\makebox(0,0)[r]{\strut{} 0.01}}%
		\put(2475,4517){\makebox(0,0)[r]{\strut{} 0.02}}%
		\put(2475,5950){\makebox(0,0)[r]{\strut{} 0.03}}%
		\put(2475,7383){\makebox(0,0)[r]{\strut{} 0.04}}%
		\put(2475,8817){\makebox(0,0)[r]{\strut{} 0.05}}%
		\put(2475,10250){\makebox(0,0)[r]{\strut{} 0.06}}%
		\put(4061,1100){\makebox(0,0){\strut{} 20}}%
		\put(5518,1100){\makebox(0,0){\strut{} 40}}%
		\put(6976,1100){\makebox(0,0){\strut{} 60}}%
		\put(8433,1100){\makebox(0,0){\strut{} 80}}%
		\put(9890,1100){\makebox(0,0){\strut{} 100}}%
		\put(11347,1100){\makebox(0,0){\strut{} 120}}%
		\put(12804,1100){\makebox(0,0){\strut{} 140}}%
		\put(14261,1100){\makebox(0,0){\strut{} 160}}%
		\put(15718,1100){\makebox(0,0){\strut{} 180}}%
		\put(17175,1100){\makebox(0,0){\strut{} 200}}%
		\put(550,5950){\rotatebox{90}{\makebox(0,0){\strut{}$\mathrm{P}(N_{\mathrm{ch}})$}}}%
		\put(9962,275){\makebox(0,0){\strut{}multiplicity}}%
		\put(14950,9675){\makebox(0,0)[r]{\strut{}PYTHIA 6.414}}%
		\put(14950,9125){\makebox(0,0)[r]{\strut{}PHOJET 1.12}}%
	\end{picture}%
\endgroup
 

%% file: plots/DiffPyt/plot_pTND-6.tex
\begin{picture}(0,0)%
	\includegraphics{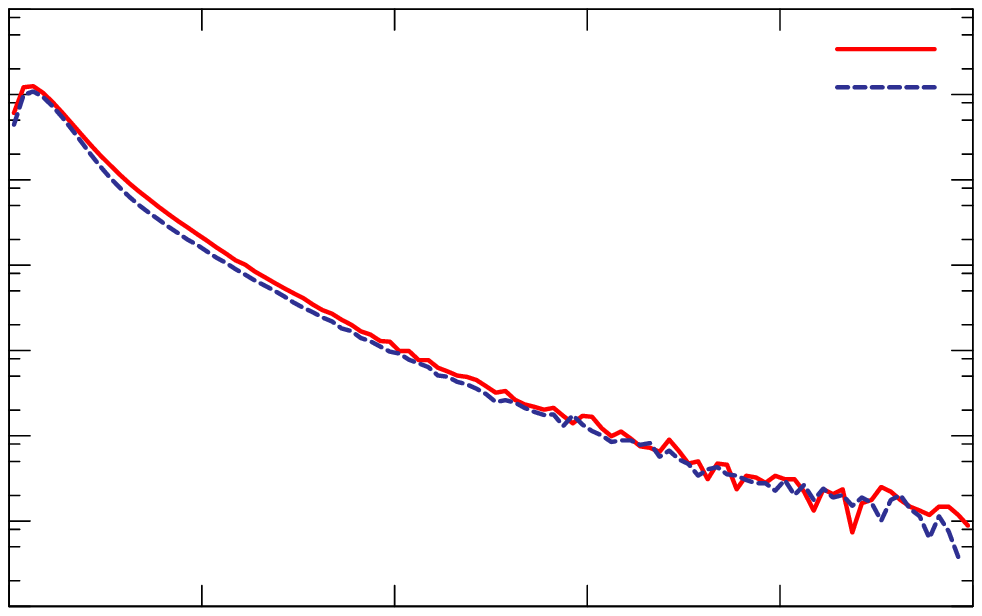}%
\end{picture}%
\begingroup
	\setlength{\unitlength}{0.0200bp}%
	\begin{picture}(18000,10800)(0,0)%
		\put(3025,1650){\makebox(0,0)[r]{\strut{} 0.0001}}%
		\put(3025,2879){\makebox(0,0)[r]{\strut{} 0.001}}%
		\put(3025,4107){\makebox(0,0)[r]{\strut{} 0.01}}%
		\put(3025,5336){\makebox(0,0)[r]{\strut{} 0.1}}%
		\put(3025,6564){\makebox(0,0)[r]{\strut{} 1}}%
		\put(3025,7793){\makebox(0,0)[r]{\strut{} 10}}%
		\put(3025,9021){\makebox(0,0)[r]{\strut{} 100}}%
		\put(3025,10250){\makebox(0,0)[r]{\strut{} 1000}}%
		\put(3300,1100){\makebox(0,0){\strut{} 0}}%
		\put(6075,1100){\makebox(0,0){\strut{} 2}}%
		\put(8850,1100){\makebox(0,0){\strut{} 4}}%
		\put(11625,1100){\makebox(0,0){\strut{} 6}}%
		\put(14400,1100){\makebox(0,0){\strut{} 8}}%
		\put(17175,1100){\makebox(0,0){\strut{} 10}}%
		\put(550,5950){\rotatebox{90}{\makebox(0,0){\strut{}$dN_{\mathrm{ch}}/dp_{\mathrm{T}}$}}}%
		\put(10237,275){\makebox(0,0){\strut{}$p_{\mathrm{T}}$ (GeV)}}%
		\put(14950,9675){\makebox(0,0)[r]{\strut{}PYTHIA 6.414}}%
		\put(14950,9125){\makebox(0,0)[r]{\strut{}PHOJET 1.12}}%
	\end{picture}%
\endgroup
 

%% file: plots/DiffPyt/plot_pTSD-6.tex
\begin{picture}(0,0)%
	\includegraphics{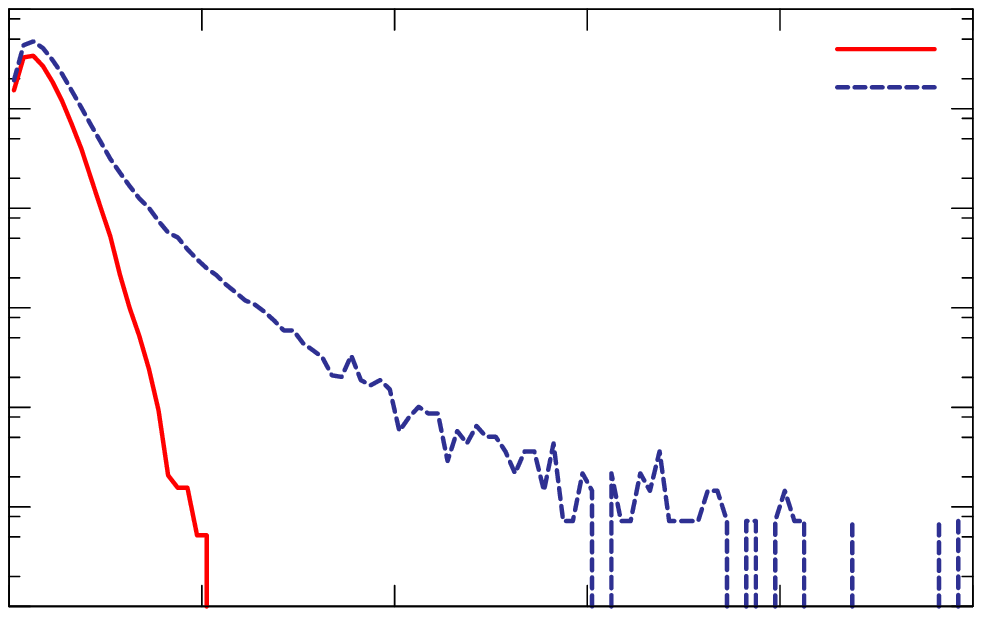}%
\end{picture}%
\begingroup
	\setlength{\unitlength}{0.0200bp}%
	\begin{picture}(18000,10800)(0,0)%
		\put(3025,1650){\makebox(0,0)[r]{\strut{} 0.0001}}%
		\put(3025,3083){\makebox(0,0)[r]{\strut{} 0.001}}%
		\put(3025,4517){\makebox(0,0)[r]{\strut{} 0.01}}%
		\put(3025,5950){\makebox(0,0)[r]{\strut{} 0.1}}%
		\put(3025,7383){\makebox(0,0)[r]{\strut{} 1}}%
		\put(3025,8817){\makebox(0,0)[r]{\strut{} 10}}%
		\put(3025,10250){\makebox(0,0)[r]{\strut{} 100}}%
		\put(3300,1100){\makebox(0,0){\strut{} 0}}%
		\put(6075,1100){\makebox(0,0){\strut{} 2}}%
		\put(8850,1100){\makebox(0,0){\strut{} 4}}%
		\put(11625,1100){\makebox(0,0){\strut{} 6}}%
		\put(14400,1100){\makebox(0,0){\strut{} 8}}%
		\put(17175,1100){\makebox(0,0){\strut{} 10}}%
		\put(550,5950){\rotatebox{90}{\makebox(0,0){\strut{}$dN_{\mathrm{ch}}/dp_{\mathrm{T}}$}}}%
		\put(10237,275){\makebox(0,0){\strut{}$p_{\mathrm{T}}$ (GeV)}}%
		\put(14950,9675){\makebox(0,0)[r]{\strut{}PYTHIA 6.414}}%
		\put(14950,9125){\makebox(0,0)[r]{\strut{}PHOJET 1.12}}%
	\end{picture}%
\endgroup
 

%% file: plots/DiffPyt/plot_lnM2.tex
\begin{picture}(0,0)%
	\includegraphics{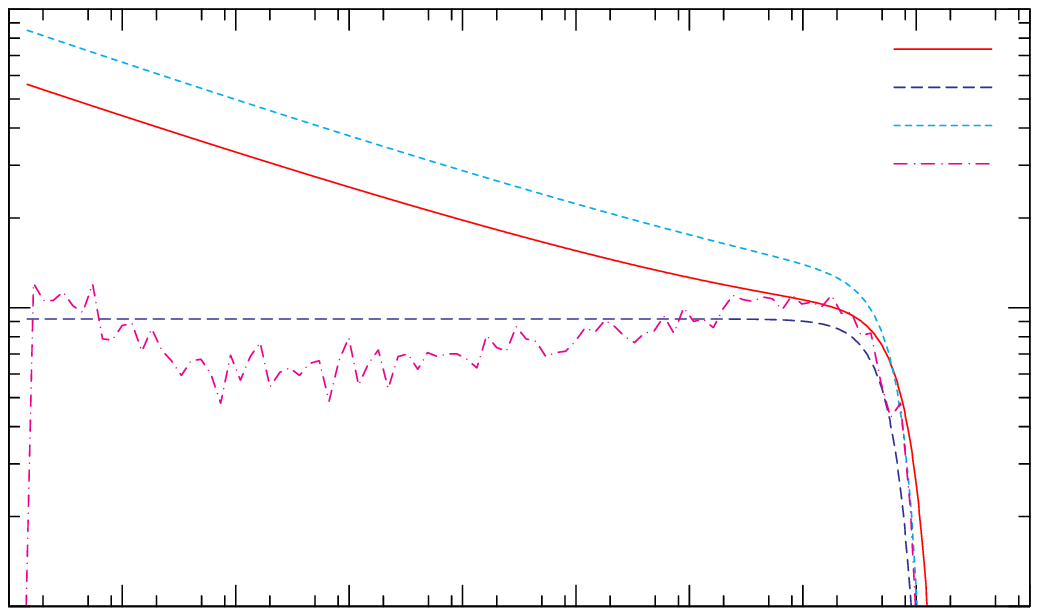}%
\end{picture}%
\begingroup
	\setlength{\unitlength}{0.0200bp}%
	\begin{picture}(18000,10800)(0,0)%
		\put(2200,1650){\makebox(0,0)[r]{\strut{} 0.1}}%
		\put(2200,5950){\makebox(0,0)[r]{\strut{} 1}}%
		\put(2200,10250){\makebox(0,0)[r]{\strut{} 10}}%
		\put(2475,1100){\makebox(0,0){\strut{} 1}}%
		\put(4108,1100){\makebox(0,0){\strut{} 10}}%
		\put(5742,1100){\makebox(0,0){\strut{} $10^2$}}%
		\put(7375,1100){\makebox(0,0){\strut{} $10^3$}}%
		\put(9008,1100){\makebox(0,0){\strut{} $10^4$}}%
		\put(10642,1100){\makebox(0,0){\strut{} $10^5$}}%
		\put(12275,1100){\makebox(0,0){\strut{} $10^6$}}%
		\put(13908,1100){\makebox(0,0){\strut{} $10^7$}}%
		\put(15542,1100){\makebox(0,0){\strut{} $10^8$}}%
		\put(17175,1100){\makebox(0,0){\strut{} $10^9$}}%
		\put(550,5950){\rotatebox{90}{\makebox(0,0){\strut{}$d\sigma_{\textrm{SD}}(\mathrm{mb})/d\ln(M^2)$}}}%
		\put(9825,275){\makebox(0,0){\strut{}$M^2(\textrm{GeV}^2)$}}%
		\put(14950,9675){\makebox(0,0)[r]{\strut{}Donnachie-Landshoff}}%
		\put(14950,9125){\makebox(0,0)[r]{\strut{}Ingelman-Bruni}}%
		\put(14950,8575){\makebox(0,0)[r]{\strut{}Streng-Berger}}%
		\put(14950,8025){\makebox(0,0)[r]{\strut{}Schuler-Sj\"{o}strand}}%
	\end{picture}%
\endgroup
 

%% file: plots/DiffPyt/plot_etaSD.tex
\begin{picture}(0,0)%
	\includegraphics{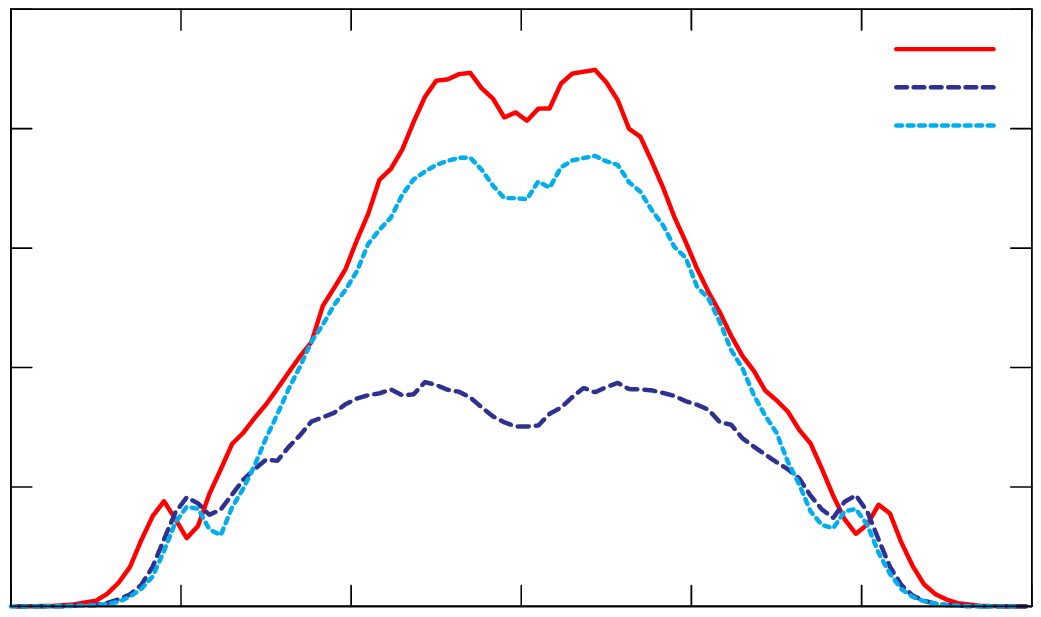}%
\end{picture}%
\begingroup
	\setlength{\unitlength}{0.0200bp}%
	\begin{picture}(18000,10800)(0,0)%
		\put(2200,1650){\makebox(0,0)[r]{\strut{} 0}}%
		\put(2200,3370){\makebox(0,0)[r]{\strut{} 0.5}}%
		\put(2200,5090){\makebox(0,0)[r]{\strut{} 1}}%
		\put(2200,6810){\makebox(0,0)[r]{\strut{} 1.5}}%
		\put(2200,8530){\makebox(0,0)[r]{\strut{} 2}}%
		\put(2200,10250){\makebox(0,0)[r]{\strut{} 2.5}}%
		\put(2475,1100){\makebox(0,0){\strut{}-15}}%
		\put(4925,1100){\makebox(0,0){\strut{}-10}}%
		\put(7375,1100){\makebox(0,0){\strut{}-5}}%
		\put(9825,1100){\makebox(0,0){\strut{} 0}}%
		\put(12275,1100){\makebox(0,0){\strut{} 5}}%
		\put(14725,1100){\makebox(0,0){\strut{} 10}}%
		\put(17175,1100){\makebox(0,0){\strut{} 15}}%
		\put(550,5950){\rotatebox{90}{\makebox(0,0){\strut{}$dN_{\mathrm{ch}}/d\eta$}}}%
		\put(9825,275){\makebox(0,0){\strut{}$\eta$}}%
		\put(14950,9675){\makebox(0,0)[r]{\strut{}PYTHIA 8.130}}%
		\put(14950,9125){\makebox(0,0)[r]{\strut{}PYTHIA 6.414}}%
		\put(14950,8575){\makebox(0,0)[r]{\strut{}PHOJET 1.12}}%
	\end{picture}%
\endgroup
 

%% file: plots/DiffPyt/plot_pTSD.tex
\begin{picture}(0,0)%
	\includegraphics{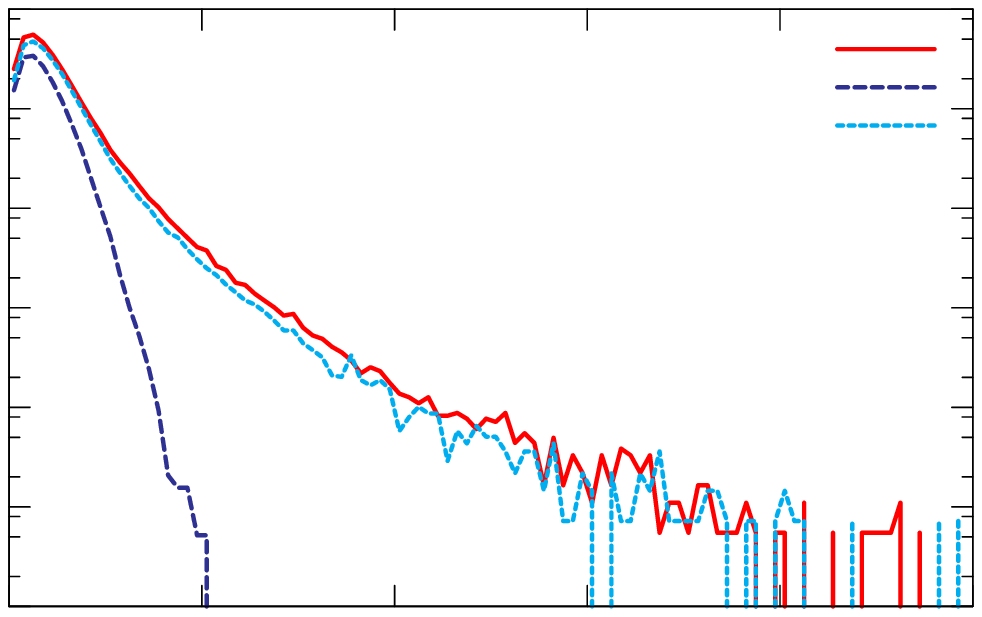}%
\end{picture}%
\begingroup
	\setlength{\unitlength}{0.0200bp}%
	\begin{picture}(18000,10800)(0,0)%
		\put(3025,1650){\makebox(0,0)[r]{\strut{} 0.0001}}%
		\put(3025,3083){\makebox(0,0)[r]{\strut{} 0.001}}%
		\put(3025,4517){\makebox(0,0)[r]{\strut{} 0.01}}%
		\put(3025,5950){\makebox(0,0)[r]{\strut{} 0.1}}%
		\put(3025,7383){\makebox(0,0)[r]{\strut{} 1}}%
		\put(3025,8817){\makebox(0,0)[r]{\strut{} 10}}%
		\put(3025,10250){\makebox(0,0)[r]{\strut{} 100}}%
		\put(3300,1100){\makebox(0,0){\strut{} 0}}%
		\put(6075,1100){\makebox(0,0){\strut{} 2}}%
		\put(8850,1100){\makebox(0,0){\strut{} 4}}%
		\put(11625,1100){\makebox(0,0){\strut{} 6}}%
		\put(14400,1100){\makebox(0,0){\strut{} 8}}%
		\put(17175,1100){\makebox(0,0){\strut{} 10}}%
		\put(550,5950){\rotatebox{90}{\makebox(0,0){\strut{}$dN_{\mathrm{ch}}/dp_{\mathrm{T}}$}}}%
		\put(10237,275){\makebox(0,0){\strut{}$p_{\mathrm{T}}$ (GeV)}}%
		\put(14950,9675){\makebox(0,0)[r]{\strut{}PYTHIA 8.130}}%
		\put(14950,9125){\makebox(0,0)[r]{\strut{}PYTHIA 6.414}}%
		\put(14950,8575){\makebox(0,0)[r]{\strut{}PHOJET 1.12}}%
	\end{picture}%
\endgroup
 

%% file: plots/DiffPyt/plot_multSD.tex
\begin{picture}(0,0)%
	\includegraphics{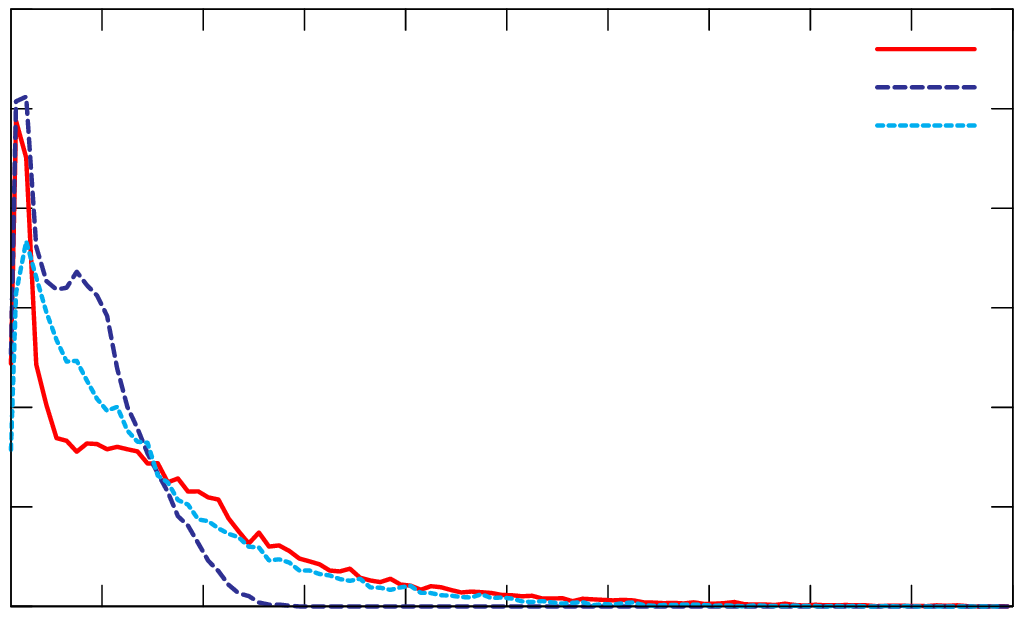}%
\end{picture}%
\begingroup
	\setlength{\unitlength}{0.0200bp}%
	\begin{picture}(18000,10800)(0,0)%
		\put(2475,1650){\makebox(0,0)[r]{\strut{} 0}}%
		\put(2475,3083){\makebox(0,0)[r]{\strut{} 0.01}}%
		\put(2475,4517){\makebox(0,0)[r]{\strut{} 0.02}}%
		\put(2475,5950){\makebox(0,0)[r]{\strut{} 0.03}}%
		\put(2475,7383){\makebox(0,0)[r]{\strut{} 0.04}}%
		\put(2475,8817){\makebox(0,0)[r]{\strut{} 0.05}}%
		\put(2475,10250){\makebox(0,0)[r]{\strut{} 0.06}}%
		\put(4061,1100){\makebox(0,0){\strut{} 20}}%
		\put(5518,1100){\makebox(0,0){\strut{} 40}}%
		\put(6976,1100){\makebox(0,0){\strut{} 60}}%
		\put(8433,1100){\makebox(0,0){\strut{} 80}}%
		\put(9890,1100){\makebox(0,0){\strut{} 100}}%
		\put(11347,1100){\makebox(0,0){\strut{} 120}}%
		\put(12804,1100){\makebox(0,0){\strut{} 140}}%
		\put(14261,1100){\makebox(0,0){\strut{} 160}}%
		\put(15718,1100){\makebox(0,0){\strut{} 180}}%
		\put(17175,1100){\makebox(0,0){\strut{} 200}}%
		\put(550,5950){\rotatebox{90}{\makebox(0,0){\strut{}$\mathrm{P}(N_{\mathrm{ch}})$}}}%
		\put(9962,275){\makebox(0,0){\strut{}multiplicity}}%
		\put(14950,9675){\makebox(0,0)[r]{\strut{}PYTHIA 8.130}}%
		\put(14950,9125){\makebox(0,0)[r]{\strut{}PYTHIA 6.414}}%
		\put(14950,8575){\makebox(0,0)[r]{\strut{}PHOJET 1.12}}%
	\end{picture}%
\endgroup
 

%% file: conclusions.tex
\section {Conclusions and future outlook}

\noindent Diffraction is not well understood, and the method employed in describing diffractive processes in PYTHIA is only one among several that have been proposed. This approach to sub-dividing the Pomeron-specific parts of the generation into independent sections may not necessarily represent the sequence of events in reality. It is important to view the effects as a convolution of factors. For example, the total diffractive cross-section is the effect of convoluting the Pomeron flux with a Pomeron-proton total cross-section. Neither the Pomeron flux, nor the total Pomeron-proton cross-section are known from first principles. This leads to a significant uncertainty in the flux factor. \\

\noindent The value of the assumed Pomeron-proton effective cross-section used in PYTHIA is \unit[10]{mb}. Increasing this value reduces the multiple interaction activity per event but if increased too much, $p_{T0}$ will be adjusted downwards to ensure that the integrated perturbative cross-section stays above the assumed total cross-section. This is the main tuneable parameter in high-mass diffraction.\\

\noindent To further complicate this picture, it is possible that an event that involves a Pomeron-proton collision that could have given a diffractive event, in addition, also involves normal multiple interactions. This would lead to a topology without rapidity gaps \cite{guillanos}. Experimentally, such events are not triggered as diffractive events.\\

\noindent A point worth mentioning is that in PYTHIA only the Schuler-Sj\"{o}strand description of the Pomeron flux includes a separate behaviour of $t$ distribution for double diffraction. Since double diffractive events do not have an outgoing proton, it is experimentally difficult to measure the $t$ distribution. \\

\noindent Central diffraction, although tiny, contributes to the total pp cross-section. Its addition to PYTHIA can be foreseen in the not so distant future.\\

\noindent With the inclusion of this perturbative description of diffraction in PYTHIA, the kinematic predictions of PYTHIA in the diffractive areas is now comparable with PHOJET, which uses a different but related physics model called the Dual Parton Model (DPM) \cite{dpm}. Although the diffractive part of PYTHIA has made considerable progress, there are still some issues that need addressing. Most importantly, comparisions with data have to be made in order to validate the model.\\

%% file: acknowledgements.tex
\section*{Acknowledgements}
\addcontentsline{toc}{section}{Acknowledgements}

\noindent This four-months project to include diffraction in PYTHIA was funded by MCnet, an European Union funded Marie Curie Research Training Network dedicated to developing the next generation of Monte Carlo event generators. I would like to thank MCnet for the opportunity and the resources.\\

\noindent I would like to thank the ALICE collaboration, especially Andreas Morsch who helped in finding the project. I would also like to thank my PhD supervisors - Roman Lietava and Cristina Lazzeroni, along with Marek Bombara, Orlando Villalobos Baillie and Paul Newman from Birmingham University for help and support.\\

\noindent A very special thanks to my supervisor during the project, Torbj\"{o}rn Sj\"{o}strand, of Lund University, who spent many hours explaining the physics and implementation of diffractive processes and who was the biggest source of help and support. The actual implementation in PYTHIA was done by him.